\documentclass[]{aa}

\usepackage{longtable}
\usepackage{natbib}
\usepackage{graphicx}
\usepackage{txfonts}

\begin{document}

\title{The Primordial Binary Population}
\subtitle{I: A near-infrared adaptive optics search for 
  close visual companions to A star members of Scorpius OB2}

 \author{M.B.N. Kouwenhoven\inst{1}
          \and
          A.G.A. Brown\inst{2}
          \and
	  H. Zinnecker\inst{3}
	  \and
	  L. Kaper\inst{1}
	  \and
	  S.F. Portegies Zwart\inst{1,4}
          }
 
 \offprints{M.B.N. Kouwenhoven \email{kouwenho@science.uva.nl}}
 
 \institute{Astronomical Institute `Anton Pannekoek', 
   University of Amsterdam, 
   Kruislaan 403, 1098 SJ Amsterdam, The Netherlands 
   \\\email{kouwenho@science.uva.nl, lexk@science.uva.nl}
   \and
   Leiden Observatory, University of Leiden, 
   P.O. Box 9513, 2300 RA
   Leiden, The Netherlands \\\email{brown@strw.leidenuniv.nl}
   \and
   Astrophysikalisches Institut Potsdam, 
   An der Sternwarte 16, D-1
   4482, Potsdam, Germany \email{hzinnecker@aip.de}
   \and
   Section Computer Science, University of Amsterdam, 
   Kruislaan 403, 1098 SJ Amsterdam, The Netherlands
   \\ \email{spz@science.uva.nl} }
 
 \date{Received \today; accepted \today}


\abstract{ 
We present the results of a near-infrared adaptive optics survey
with the aim to detect close companions to {\it Hipparcos} members in the
three subgroups of the nearby OB~association Sco~OB2: Upper Scorpius (US),
Upper Centaurus Lupus (UCL) and Lower Centaurus Crux (LCC). We have targeted
199 A-type and late B-type stars in the $K_S$ band, and a subset also in the
$J$ and $H$ band. We find 151 stellar components other than the target
stars. A brightness criterion is used to separate these components into 77
background stars and 74 candidate physical companion stars. 
Out of these 74 candidate companions, 41 have not
been reported before (14 in US; 13 in UCL; 14 in LCC). 
The angular separation between primaries and observed companion stars 
ranges from $0.22''$ to $12.4''$. At the mean distance of Sco~OB2 (130~pc) 
this corresponds to a projected separation of $28.6$~AU to $1612$~AU. 
Absolute magnitudes are derived for all primaries and observed companions
using the parallax and interstellar extinction for each star individually.
For each object we derive the mass from $K_S$, assuming an age of 5~Myr for the
US subgroup,
and 20~Myr for the UCL and LCC subgroups.
Companion star masses range 
from $0.10~{\rm M}_\odot$ to $3.0~{\rm M}_\odot$.
The mass ratio distribution follows $f(q) = q^{-\Gamma}$ with $\Gamma=0.33$, 
which excludes random pairing.
No close ($\rho \leq 3.75''$) companion stars or background stars 
are found in the magnitude range $12~{\rm mag}\leq K_S \leq 14~{\rm mag}$.
The lack of stars with these properties
cannot be explained by low-number statistics, and may imply a lower
limit on the companion mass of $\sim 0.1~{\rm M}_\odot$. 
Close stellar components with $K_S>14~{\rm mag}$ are observed. If these
components are very low-mass companion stars, a gap
in the companion mass distribution might be present.
The small number of close low-mass companion stars
could support the embryo-ejection formation scenario for brown dwarfs.
Our findings are compared
with and complementary to visual, spectroscopic, and astrometric data on
binarity in Sco~OB2. 
We find an overall companion star fraction of 0.52 in this association. 
This is a lower limit since the data from the observations and
from literature are
hampered by observational biases and selection effects.
This paper is the first step toward our goal 
to derive the primordial binary population in Sco~OB2.

\keywords{binaries: visual -- binaries: general -- stars: formation -- associations -- individual: Sco~OB2}

 }

\maketitle


\section{Introduction}

Duplicity and multiplicity 
properties of newly born stars are among the most important clues
to understanding the process of star formation \citep{blaauw1991}.
Observations of star forming regions over the past two decades have revealed
two important facts: (1) practically all (70--90\%) stars form in clusters 
\citep[e.g.,][]{LL2003} and, (2) within these clusters most stars are 
formed in binaries \citep{mathieu1994}.
Consequently, the star formation community has
shifted its attention toward understanding the formation of multiple systems
--- from binaries to star clusters --- by means of both observations and
theory.

The observational progress in studies of very young embedded as well as
exposed star clusters is extensively summarized in the review by
\cite{LL2003}. On the theoretical side the numerical simulations of cluster
formation have become increasingly sophisticated, covering the very earliest
phases of the development of massive dense cores in giant molecular clouds
\citep[e.g.,][]{klessen2000}, 
the subsequent clustered formation of
stars and binaries \citep[e.g.,][]{bate2003}, 
as well as the early evolution of the binary population
during the phase of gas expulsion from the
newly formed cluster \citep{kroupa2001}.
At the same time numerical
simulations of older exposed clusters have become more realistic
by incorporating detailed stellar and binary evolution effects in N-body
simulations \citep[e.g.,][]{spz2001}. This has led to the creation of a number
of research networks that aim at synthesizing the modeling and observing
efforts into a single framework which covers all the stages from the formation
of a star cluster to its eventual dissolution into the Galactic field, an
example of which is the MODEST collaboration \citep{hut2003,sills2003}.

Such detailed models require stringent observational constraints in the form
of a precise characterization of the stellar content of young
clusters. Investigations of the stellar population in young clusters have
mostly focused on single stars. However, as pointed out by \cite{larson2001},
single stars only retain their mass from the time of formation whereas
binaries retain three additional parameters, their mass ratio, angular
momentum and eccentricity. Thus, the properties of the binary population can
place much stronger constraints on the physical mechanisms underlying the star
and cluster formation process.

Ideally, one would like an accurate description of the `primordial' binary
population. This population was defined by \cite{brown2001} as ``the
population of binaries as established just after the formation process itself
has finished, i.e., after the stars have stopped accreting gas from their
surroundings, but before stellar or dynamical evolution have had a chance to
alter the distribution of binary parameters''. This definition is not entirely
satisfactory as stellar and dynamical evolution will take place already during
the gas-accretion phase.

Here we revise this definition to: 
{\em``the population of binaries as established
just after the gas has been removed from the forming system, i.e., when the
stars can no longer accrete gas from their surroundings''}.
This refers to the
same point in time, but the interpretation of the `primordial binary
population' is somewhat different. The term now refers to the point in time
beyond which the freshly formed binary population is affected by {\em
stellar/binary evolution and stellar dynamical effects only}. Interactions with a surrounding gaseous medium no longer take place.

From the point of view of theoretical/numerical models of star cluster
formation and evolution the primordial binary population takes on the
following meaning. It is the final population predicted by simulations of the
formation of binaries and star clusters, and it is the initial population for
simulations that follow the evolution of star clusters and take into account
the details of stellar dynamics and star and binary evolution. 
The primordial
binary population as defined in this paper 
can be identified with the initial binary population,
defined by \cite{kroupa1995b} as the binary population at the 
instant in time when the pre-main-sequence eigenevolution
has ceased, 
and when dynamical evolution of the stellar cluster becomes effective.
\cite{kroupa1995a,kroupa1995c} infers the
initial binary population by the so-called inverse dynamical population
synthesis technique.
This method involves the evolution of simulated stellar
clusters forward in time for different initial binary populations, where the
simulations are repeated until a satisfactory fit with the present day binary
population is found.

Our aim is to obtain a detailed observational characterization of the
primordial binary population as a function of stellar mass, binary parameters,
and (star forming) environment. The most likely sites where this population
can be found are very young (i.e., freshly exposed), 
low density stellar groupings containing a
wide spectrum of stellar masses. 
The youth of such a stellar grouping implies that stellar
evolution will have affected the binary parameters of only a handful of the
most massive systems.
The low stellar densities guarantee that
little dynamical evolution has taken place
after the gas has been removed from the forming system.
These constraints naturally
lead to the study of the local ensemble of OB associations. 
Star clusters are older and have a higher density than OB associations
and are therefore less favorable.
For example, in the Hyades and Pleiades,
the binary population has significantly changed 
due to dynamical and stellar evolution \citep{kroupa1995c}.
Note that OB associations may start out as dense clusters 
\citep{kroupa2001,kroupaboily2002}.
However, they rapidly expel their gas and evolve to low-density systems. 
This halts any further dynamical evolution of the binary population.
\cite{brown1999} define OB associations as 
``young ($\la$\,50~Myr) stellar groupings
of low density ($\lesssim 0.1 {\rm M}_\odot {\rm pc}^{-3}$) 
---~such that they are likely to be unbound~--- 
containing a significant population of B stars.''
Their projected dimensions range from
$\sim$\,10 to $\sim$\,100~pc and their mass spectra cover the mass range from
O~stars all the way down to brown dwarfs. 
For reviews on OB
associations we refer to \cite{blaauw1991} and \cite{brown1999}.  Thanks to
the {\it Hipparcos} Catalogue (ESA 1997) the stellar content of the nearby OB
associations has been established with
unprecedented accuracy to a completeness limit of $V\,\sim\,10.5~{\rm mag}$, or about
1~M$_\odot$ for the stars in the nearest associations 
\citep[][]{dezeeuw1999, hoogerwerf2000}. Beyond this limit the population of
low-mass pre-main-sequence stars has been intensively studied in, e.g., the
Sco~OB2 association \citep[][]{preibisch2002,mamajek2002}.

The latter is also the closest and best studied of the OB associations in the
solar vicinity and has been the most extensively surveyed for binaries. 
The association consists of three
subgroups: Upper Scorpius (US, near the Ophiuchus star forming region, at a
distance of 145\,pc), Upper Centaurus Lupus (UCL, 140\,pc) and Lower Centaurus
Crux (LCC, 118\,pc). The ages of the subgroups range from 5 to $\sim20$\,Myr
and their stellar content has been established from OB stars down to brown
dwarfs.
\citep[for details see][]{degeus1989, dezeeuw1999, debruijne1999,
hoogerwerf2000, mamajek2002, preibisch2002}. 
Surveys targeting the binary
population of Sco~OB2 include the radial-velocity study by \cite{levato1987},
the speckle interferometry study by \cite{koehler2000}, and 
the adaptive optics study by 
\cite{shatsky2002}. Because of their brightness many of the B-star members of
Sco OB2 have been included in numerous binary star surveys, in which a
variety of techniques have been employed. The literature data on the binary
population in Sco~OB2 is discussed by \cite{brown2001} and reveals that
between 30 and 40 per cent of the {\it Hipparcos} members of Sco~OB2 are known to be
binary or multiple systems. However, these data are incomplete and suffer from
severe selection effects, which, if not properly understood, will prevent a
meaningful interpretation of the multiplicity data for this association in
terms of the primordial binary population. The first problem can be addressed
by additional multiplicity surveys of Sco~OB2. In this paper we report on
our adaptive optics survey of Sco~OB2 which was aimed at surveying all the
{\it Hipparcos} members of spectral type A and late B, 
using the ADONIS instrument on the
3.6m telescope at ESO, La Silla.

We begin by describing in Sect.~\ref{sec:theaosurvey} our observations, the
data reduction procedures and how stellar components other than the target
stars were detected in our images. These components have to be separated into
background stars and candidate physical companions. We describe how this was
done in Sect.~\ref{sec: backgroundstars}. The properties of the physical
companions are described in Sect.~\ref{sec: properties}. In Sect.~\ref{sec:
literaturedata} we discuss which of the physical companions are new by
comparing our observations to data in the literature and we provide updated
statistics of the binary population in Sco~OB2. We summarize this work in
Sect.~\ref{sec:conclusions} and outline the next steps of this study which are
aimed at addressing in detail the problem of selection biases associated with
multiplicity surveys and subsequently characterizing the primordial binary
population.



\section{Observations and data reduction}
\label{sec:theaosurvey}

\subsection{Definition of the sample} \label{sec: sample}

A census of the stellar content of the three subgroups of Sco~OB2 based on
positions, proper motions, and parallaxes was presented by \cite{dezeeuw1999}.
Our sample is extracted from their list of \textit{Hipparcos} member stars and
consists of A and B stars. In order to avoid saturating the detector
during the observations, we were restricted to observing stars fainter than
$V\sim6~{\rm mag}$. The sample therefore consists mainly of late-B and A stars which at
the mean distance of Sco~OB2 (130 pc) corresponds to stars with $6~{\rm mag}\lesssim
V\lesssim 9~{\rm mag}$, which translates to very similar limits in $K_S$. We observed
199 stars from the resulting 
\textit{Hipparcos} member sample (Table~\ref{table: results}). 
Not all late-B and A member stars could be observed due to time limitations.
The distribution of target stars over spectral type is: 83~(157)~B, 113~(157)~A,
2~(138)~F, and 1~(48)~G, where the numbers in brackets denote the total
number of Sco~OB2 {\it Hipparcos} members of the corresponding spectral type.

\subsection{Observations} \label{sec: observations}

The observations were performed with the ADONIS/SHARPII+ system 
\citep{beuzit1997} on the ESO
3.6~meter telescope at La Silla, Chile. This is an adaptive optics system
coupled to an infrared camera with a NICMOS3 detector array. The field of
view of the $256 \times 256$ pixel detector 
is $12.76'' \times 12.76''$. The plate scale
is $0.0495'' \pm 0.0003''$ per pixel, and the orientation of the
field is $-0^\circ.20 \pm 0^\circ.29$ (measured from North to
East)\footnote{The astrometric calibrations are based on observations of the
  astrometric reference field ($\theta$~Ori) for ADONIS/SHARPII+/NICMOS3 (see
  \texttt{http://www.imcce.fr/priam/adonis}). Mean values for the period
  9/1999 to 12/2001 are used. Plate scale and position angle differences
  between the two observing runs fall within the error bars.}. 
Wavefront sensing was performed directly on the target stars.

\begin{figure}[bt]
  \centering
  \includegraphics[width=0.5\textwidth,height=!]{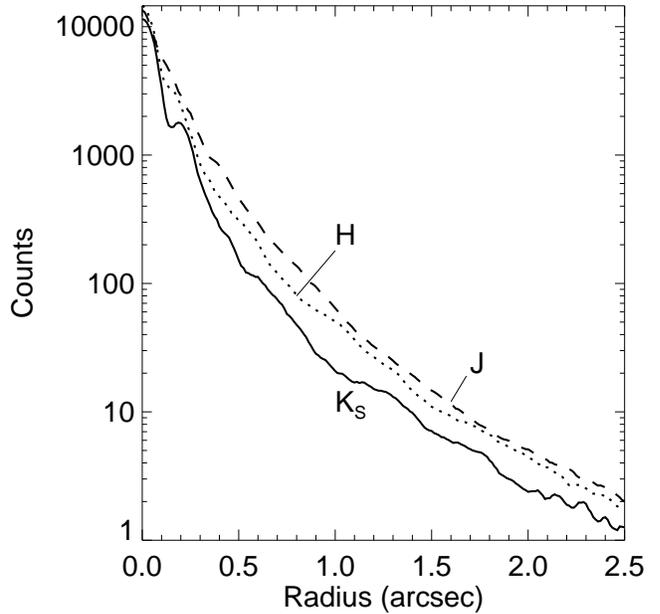}
  \caption{The radial profile of the PSF for the target star
    HIP53701. The observations are obtained using the 
    ADONIS/SHARPII+ system on June 7, 2001 in $J$, $H$, and $K_S$.
    The corresponding Strehl ratios for these observations
    are 18\% in $J$, 26\% in $H$
    and 31\% in $K_S$.
  \label{figure: psfprofile}}
\end{figure}

Our observing campaign was carried out in two periods: a period from May 31 to
June 4, 2000, and one from June 5 to 9, 2001. Out of the eight observing
nights about 1.5 were lost due to a combination of bad weather and technical
problems. Each target star is observed in the $K_S$ ($2.154~\mu$m) band. A
subset is additionally observed in the $J$ ($1.253~\mu$m) and $H$
($1.643~\mu$m) band. In our search for companion stars, near-infrared
observations offer an advantage over observations in the optical. In the
near-infrared the luminosity contrast between the primary star and its (often
later type) companion(s) is lower, which facilitates the detection of faint
companions. The performance of the AO system is measured by the Strehl ratio, 
which is the ratio between the maximum of the ideal point spread
function (PSF) of the system and the measured maximum of the PSF. Typical
values for the Strehl ratio in our observations were 5--15\% in $J$, 15--20\% in $H$ and
20--35\% in $K_S$ (Figure~\ref{figure: psfprofile}).

\begin{figure}[bt]
  \centering
  \includegraphics[width=0.5\textwidth,height=!]{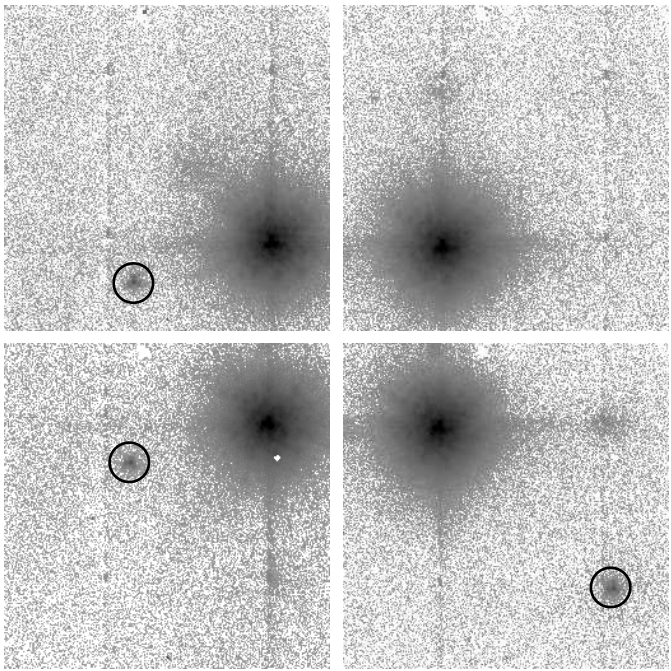}
  \caption{An example illustrating the observing strategy.
  Observations are obtained with ADONIS/SHARPII+ system in $J$, $H$, 
  and $K_S$.
  Four observations of each star are made, each observed with a
  different offset, in order to maximize the field of view. The frame size
  of the individual quadrants is $12.76'' \times 12.76''$. Combination of the quadrants
  results in an effective field of view of $19.1'' \times 19.1''$.
  This figure shows $K_S$ band
  observations of the A0~V star HIP75957. Two stellar components at
  angular separations of $5.6''$ and $9.2''$ 
  (indicated with the black circles) 
  are detected.
  These components are visible to the left (East)
  and bottom-right (South-West) of HIP75957 and are probably background 
  stars. All other features are artifacts.
  \label{figure: mosaic}}
\end{figure}

Each star was observed at four complementary pointings in order to enhance the
sensitivity of the search for close companions and to maximize the available
field of view (Figure~\ref{figure: mosaic}). The two components of the known
and relatively wide binaries HIP77315/HIP77315 and HIP80324 
were observed individually
and combined afterwards. For all targets each observation consists of 4 sets of
30 frames (i.e. 4 data cubes). The integration time for each frame was
$200-2000$ ms, depending on the brightness of the source. After each
observation, thirty sky frames with integration times equal to those of the
corresponding target star were taken in order to measure background
emission. The sky frames were taken 10 arcmin away from the target star and
care was taken that no bright star was present in the sky frame. Nevertheless,
several sky frames contain faint background stars. The error on the target
star flux determination due to these background stars is less than $0.06\%$ of
the target star flux and therefore negligible for our purposes. Dark frames
and flatfield exposures (using the internal lamp) were taken and standard
stars were regularly observed.

We did not use the coronograph in our setup. The coronograph obscures most of
the target star's flux, which complicates flux calibration of the target stars
and their companions. Furthermore, the use of the coronograph prevents
detection of close companions with angular separations
less than $1''$. The advantage of the
coronograph is that one can detect fainter objects, which would otherwise
remain undetected due to saturation of the target stars. However, the large
majority of these faint components are likely background stars (see
\S\ref{sec: backgroundstars}).

\subsection{Data reduction procedures} \label{sec: datareduction}

The primary data reduction was performed with the ECLIPSE package
\citep{eclipseref}. Information about the sensitivity of each pixel is derived
from the (dark current subtracted) flatfield images with different exposure
times. The number of counts as a function of integration time is linearly
fitted for each pixel. The results are linear gain maps (pixel sensitivity
maps) for each filter and observing night. Bad pixel maps for each observing
run and filter are derived from the linear gain maps: each pixel with a
deviation larger than $3\sigma$ from the median is flagged as a bad pixel. The
average sky map is subtracted from the corresponding standard star and target
star data cubes. The resulting data cubes are then corrected using the dark
current maps, the linear gain maps, and bad pixel maps.

Image selection and image combination was done with software that was written
specifically for this purpose. 
Before combination, frames with low Strehl ratios with respect to the
median were removed.  Saturated frames (i.e. where the stellar flux
exceeds the SHARPII+ linearity limit) and frames with severe wavefront
correction errors were also removed. 
The other frames were combined by taking the median.
Typically, about 2 to 3 out of 30 frames were rejected
before combination.

\subsection{Component detection} \label{sec: componentdetection}

We used the STARFINDER package \citep{diolaiti2000} to determine the position
and instrumental flux of all objects in the images. For each component other
than the target star we additionally measured the correlation between the PSF
of the target star and that of the component. Components with peak fluxes less
than 2 to 3 times the noise in the data, or with correlation coefficients less
than $\sim 0.7$, were considered as spurious detections.

Since the PSF of the target star is generally not smooth, it was sometimes
difficult to decide whether a speckle in the PSF halo of the star is a stellar
component, or merely a part of the PSF structure. In this case four
diagnostics were used to discriminate between stellar components and PSF
artifacts. First, a comparison between the images with the target star located
in four different quadrants was made. Objects that do not appear in all
quadrants where they {\it could} be detected were considered
artifacts. Second, a similar comparison was done with the individual
uncombined raw data frames. Third, a radial profile was fit to the PSF and
subtracted from the image to increase the contrast, so that the stellar
component becomes more obvious. And finally, a comparison between the PSF of
the target star and the PSF of the star previous or next in the program with a
similar position on the sky was made by blinking the two images. These two
PSFs are expected to be similar and therefore faint close companion stars can
be detected.

\begin{figure}[bt]
  \includegraphics[width=0.5\textwidth,height=!]{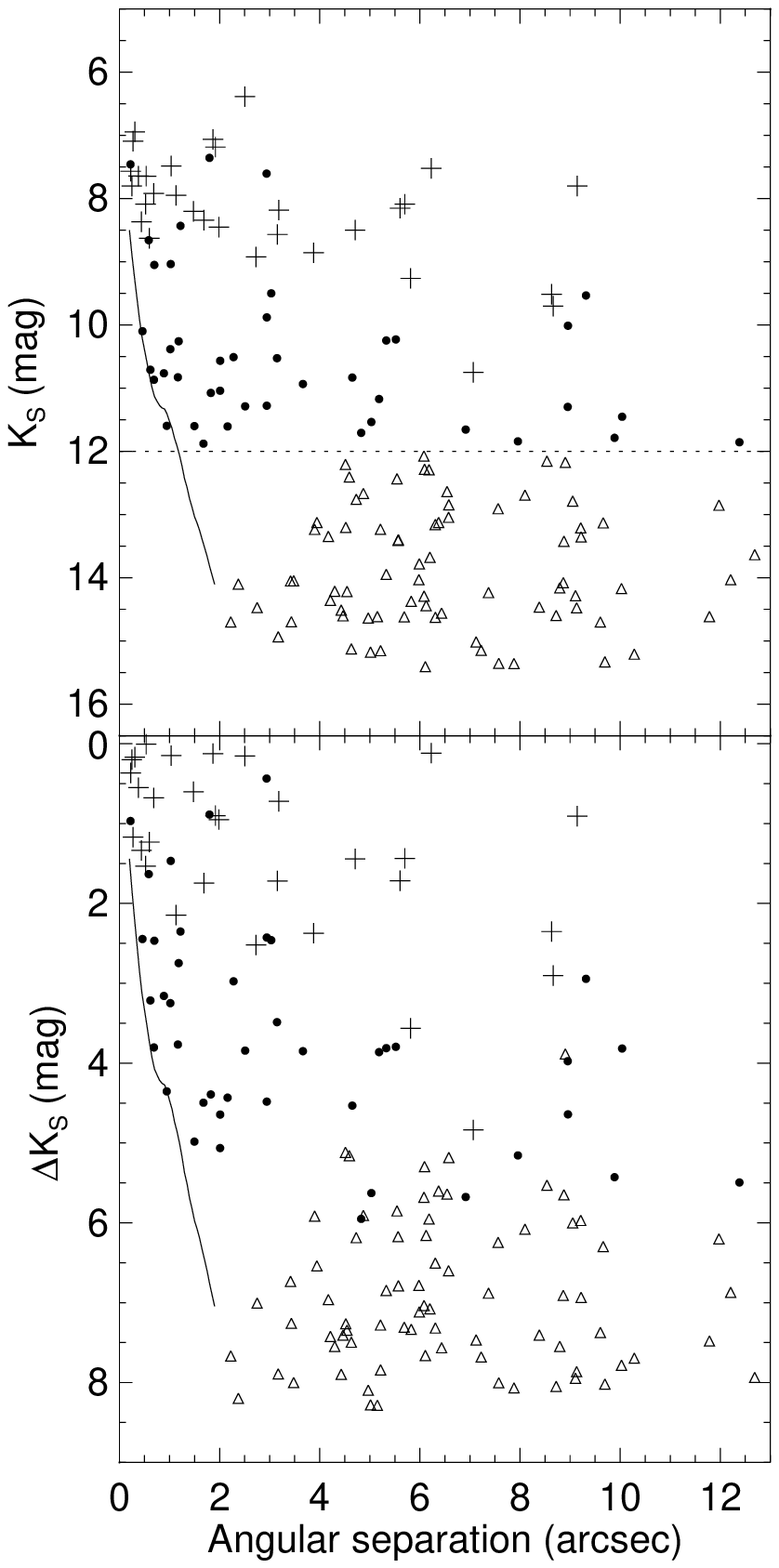}   
  \caption{Angular separation between the target stars and observed components as a
  function of the component's $K_S$ magnitude ({\it top}) and component's
  $K_S$ magnitude relative to that of the target star ({\it bottom}). 
  The figures show 
  the 41 new (dots) and 33 previously known (plusses) 
  companion stars ($K_S < 12~{\rm mag}$),
  respectively.
  The triangles are the 77 background stars 
  ($K_S > 12~{\rm mag}$; \S~\ref{sec: backgroundstars}). The solid curves show the
  estimated detection limit as a function of angular separation. 
  These
  figures clearly demonstrate
  that faint close companions are not detected because of
  the extended PSF halo of the target star. 
  The effective field of view is $19.1'' \times 19.1''$.
  Companion star stars with angular separations larger than
  $\frac{1}{2}\sqrt{2} \cdot 19.1'' = 13.5''$ cannot be detected.
  For small
  angular separation ($\rho < 3.5''$) no faint ($K_S > 12~{\rm mag}$) components
  are found. This could indicate a gap in the companion mass function, or a
  lower mass limit for companion stars at small separation 
  (see \S~\ref{sec: masses}). 
  \label{figure: detectionlimits}}
\end{figure}

In total we detect 151 stellar components other than the target stars.
A significant fraction of these 151 components are background
stars (see \S~\ref{sec: backgroundstars}). All other components
are companion stars.
Figure~\ref{figure: detectionlimits} shows the $K_S$ magnitude of
the companion and background stars as
a function of angular separation $\rho$ between the target star
and the companion or background stars. The same plot also shows
$\Delta K_S = K_{S,{\rm component}}-K_{S,{\rm primary}}$ as a
function of $\rho$. Previously unknown companion stars
(see \S~\ref{sec: newcompanions}) and known companions 
are indicated with the dots and 
the plusses, respectively. The background stars
(see \S~\ref{sec: backgroundstars}) are represented with
triangles. This figure shows that the detection probability of
companion stars increases with increasing angular separation and decreasing
magnitude difference, as expected. The lower magnitude limit for component
detection is determined by the background noise; the lower limit for the
angular separation is set by the PSF delivered by the AO system. Note that not
all detected components are necessarily companion stars. A simple criterion is
used to separate the companion stars ($K_S < 12~{\rm mag}$) and the background stars
($K_S > 12~{\rm mag}$). See \S~\ref{sec: backgroundstars} for the motivation of this
choice.

The estimated detection limit as a function of angular separation and
companion star flux is represented in Figure~\ref{figure: detectionlimits} with a
solid curve. The detection limit is based on $K_S$ band measurements of
simulated companion stars around the B9~V target star HIP65178. In the
image we artificially added a second component, by scaling and shifting a copy
of the original PSF, thus varying the desired flux and angular separation of
the simulated companion. We sampled 30 values for the companion flux linearly
in the range $9.35~{\rm mag} \leq K_S \leq 15.11~{\rm mag}$, 
and 40 values for the angular
separation between primary and companion linearly in the range $0.05''
\leq \rho \leq 2''$. For each value of the angular separation we determined
the minimum flux for which the companion star is detectable (i.e. with a peak
flux of $2-3$ times the noise in the data). In order to minimize biases, we
repeated this procedure for four different position angles and averaged the
results. The standard deviation of the detection limit is $0.07''$ at a
given magnitude and $0.35~{\rm mag}$ at a given angular separation.  Similar
simulations with other target stars (e.g., HIP58859, Strehl 
ratio = 8\%; HIP76048, Strehl ratio =
27\%) show that the detection limit for the observations with different Strehl ratios
is comparable to that of HIP65178 (Strehl ratio = 40\%). All target stars are of
spectral type A or late-B and have comparable distances. The detection limit
shown in Figure~\ref{figure: detectionlimits} is therefore representative for
our sample.

\begin{figure}[bt]
  \includegraphics[width=0.5\textwidth,height=!]{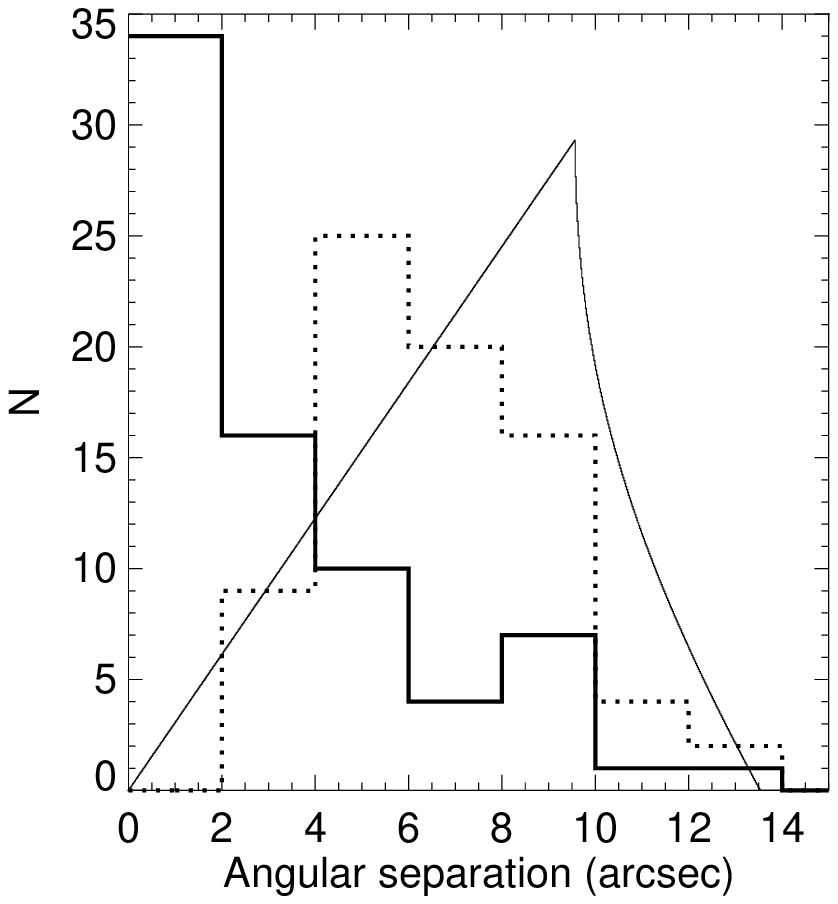} 
  \caption{The number of objects per angular separation bin versus 
  angular separation (bin size $2''$). The
  distribution of presumed background stars (dotted histogram) is clearly
  different from that of the companion stars (solid histogram). The expected
  angular separation distribution 
  for background stars (in units of stars per bin size) is represented 
  with the solid curve. The curve is normalized such that the
  number of stars corresponds to the expected number of background stars
  with $12~{\rm mag}<K_S<16~{\rm mag}$ 
  (cf. Figure~\ref{figure: expectedbackground}).
  The excess of close ($\rho \lesssim 6''$) companions
  could be caused by faint companion stars that are misclassified
  as background stars. 
  Objects with $\rho < 0.1''$ are too deep in the PSF halo of the
  primary to be detectable. Objects with $\rho > 15''$ cannot be
  measured due to the limited field of view of the ADONIS/SHARPII+ system. 
  \label{figure: separationhisto} }
\end{figure}

\subsection{Photometry}

Near-infrared magnitudes for the observed standard stars are taken from
\cite{bliek1996}, \cite{carter1995}, and the 2MASS catalog. All magnitudes are
converted to the 2MASS system using the transformation formulae of
\cite{carpenter2001}. The standard stars HD101452, HD190285 and HD96654 were
found to be double. The fluxes for the primary and secondary of the standard
stars HD101452 and HD96654 are added for calibration, since the secondaries
are unresolved in \cite{bliek1996}. For HD190285, only the flux of the primary
is used for calibration since the secondary is resolved in the 2MASS catalog.

The zeropoint magnitudes are calculated for each observing night and each
filter individually. There are non-negligible photometric zeropoint differences
between the four detector quadrants. Application of the reduction procedure to simulated
observations (but using the ADONIS/SHARPII+ dark frames and flatfield exposures) 
also shows this zeropoint offset, unless the flatfielding was
omitted. Hence, the zeropoint offset between the four quadrants is caused by
large-scale variations in the flatfield exposures. The four detector quadrants
were consequently calibrated independently. 

Since the distribution of observed standard stars over airmass is not
sufficient to solve for the extinction coefficient $k$, we use the mean
extinction coefficients for La~Silla\footnote{see
\texttt{http://www.ls.eso.org/lasilla/sciops/ntt/sofi}}: 
$k_J=0.081~{\rm mag/airmass}$, $k_H=0.058~{\rm mag/airmass}$ and
$k_{K_S}=0.113~{\rm mag/airmass}$. The standard stars
have similar spectral type as the target stars, which allows us to neglect the
color term in the photometric solution.


\section{Background stars} \label{sec: backgroundstars}

\begin{figure}[bt]
  \includegraphics[width=0.5\textwidth,height=!]{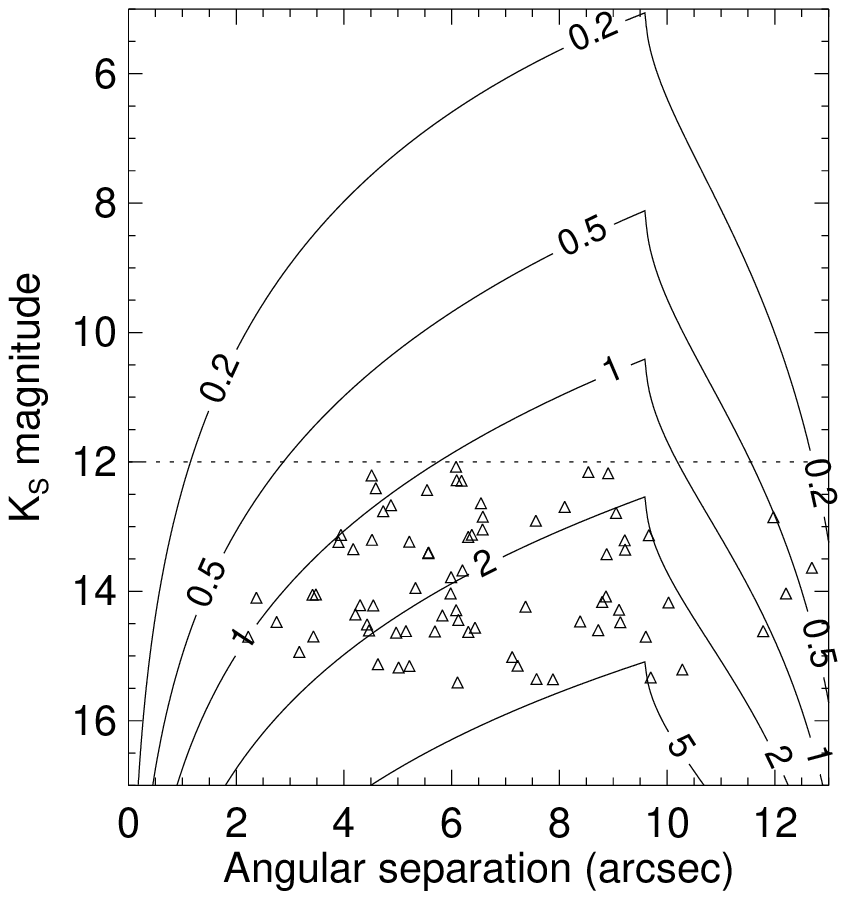}
  \caption{The expected background star distribution 
    as a function of angular separation
    and the $K_S$ magnitude distribution
    of background stars as observed by \cite{shatsky2002}. 
    The distribution of
    background stars over the field of view is assumed to be homogeneous.
    The triangles represent the observed stellar objects that we
    classify as background stars. 
    The contours indicate the background star density in units of
    arcsec$^{-1}$mag$^{-1}$. The distribution is normalized, so that for $14~{\rm mag}
    \leq K_S \leq 15~{\rm mag}$ and $5'' \leq \rho \leq 13''$ the number of
    background stars equals 19, i.e. the observed number of background
    stars with these properties. Observational biases are not considered. 
    The dashed line represents our criterion to separate
    companion stars ($K_S \leq 12~{\rm mag}$) and background stars 
    ($K_S>12~{\rm mag}$). 
    The excess of background stars at small angular separations 
    could be caused by faint companion stars that are misclassified
    as background stars.
    \label{figure: expectedbackground}}
\end{figure}

\begin{figure}[bt]
  \includegraphics[width=0.5\textwidth,height=!]{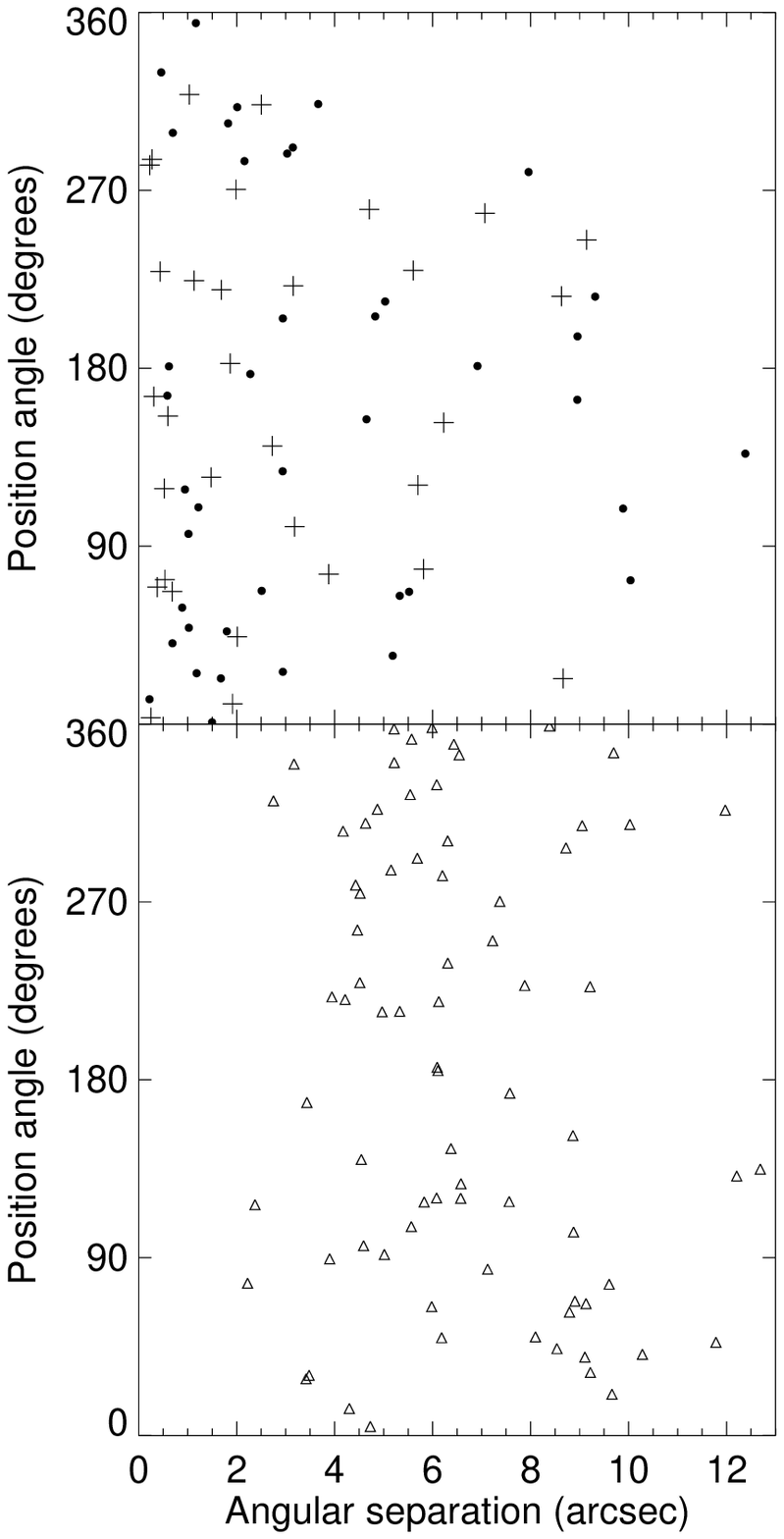} 
  \caption{ {\it Top:} the distribution of presumed companion stars 
    over angular separation and position angle.
    Previously known companions and new companions are indicated with the plusses and 
    dots, respectively.
    {\it Bottom:} the distribution of background stars 
    over angular separation and position angle.
    The distribution of companion stars and background stars over position angle
    is random for angular separation smaller than $9.6''$ (see text).
    The companion stars are more 
    centrally concentrated than the background stars.
    \label{figure: compsfield} }
\end{figure}

Not all of the 151 stellar components that we find around the target
stars are companion stars. 
The chance that a foreground or background star is present in the field is not
negligible. There are several techniques that can be applied to determine
whether a detected component is a background star or a probable companion
star.

Separation of the objects based on the position of the component in the color-magnitude
diagram (CMD) is often applied to discriminate background stars from companion
stars \citep[see, e.g., ][]{shatsky2002}. Given a certain age of the
population, a companion star should be found close to the isochrone that
corresponds to that age. When age and distance spread are properly taken into
account, this method provides an accurate estimate of the status of the
object. We are unable to apply this technique to our data since
only a small fraction of our observations is multi-color (see \S~\ref{sec:
hrdiagram} and Figure~\ref{figure: 3hr}).

\begin{figure*}[bt]
  \includegraphics[width=\textwidth,height=!]{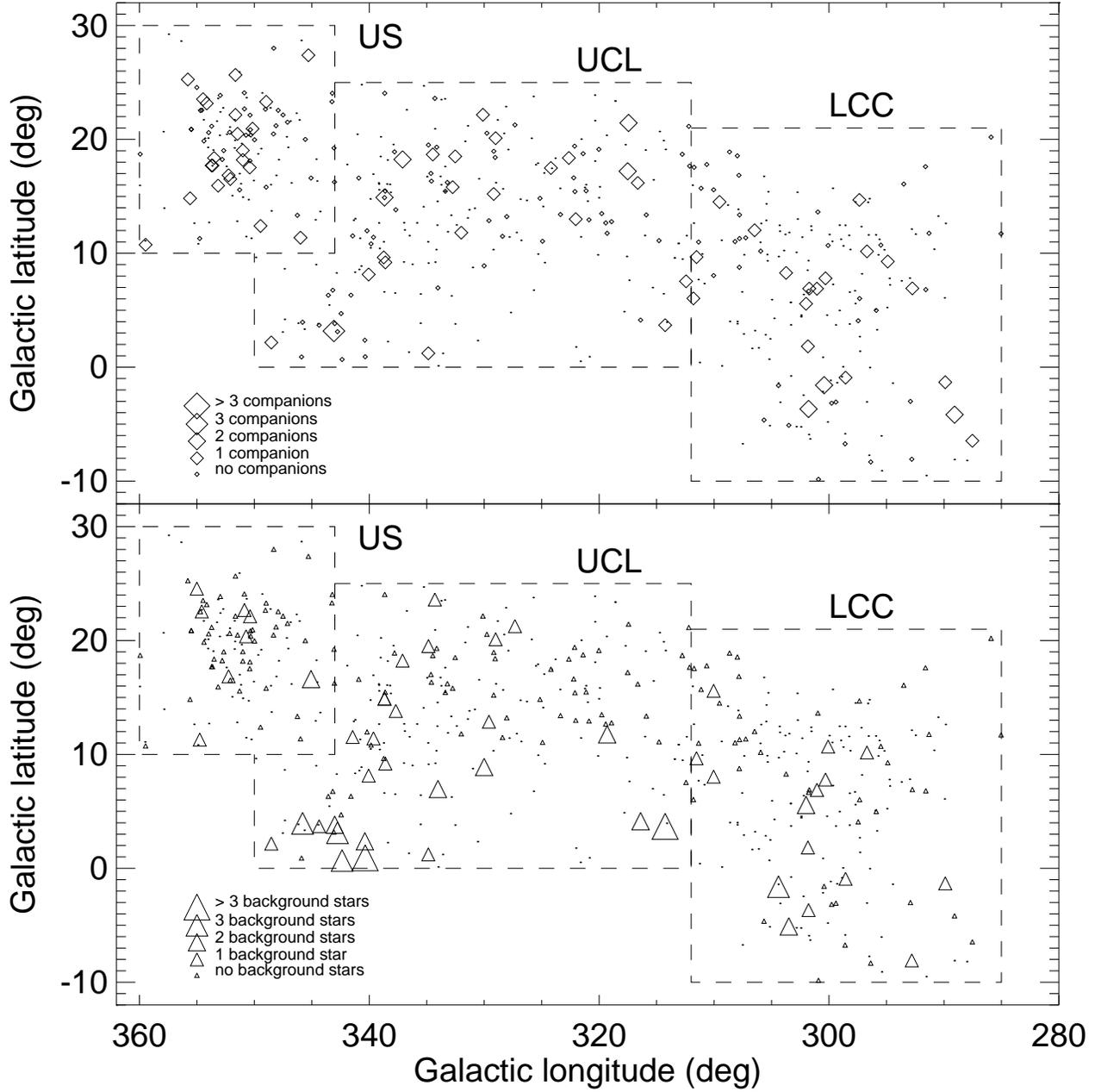}  
  \caption{The distribution of 74 detected candidate companions ({\it top}) and
  the 77 presumed background stars ({\it bottom}) over
  Galactic coordinates. The number of companions and background stars found
  per target star is indicated by the size of the diamonds and triangles,
  respectively. {\it Hipparcos} member stars that were not observed are shown
  as dots. The borders of the three subgroups of Sco~OB2 (Upper Scorpius,
  Upper Centaurus Lupus, and Lower Centaurus Crux) are indicated by the dashed
  lines. The background stars are concentrated towards the location of 
  the Galactic plane (see also Figure~\ref{figure: galacticplanehisto}).
  \label{figure: galacticplane}}
\end{figure*}

Following \cite{shatsky2002} we use a simple brightness criterion
to separate the companion stars ($K_S \leq 12~{\rm mag}$) and the background stars ($K_S >
12~{\rm mag}$). At the typical distance of a Sco~OB2, an M5 main sequence star
has approximately $K_S=12~{\rm mag}$.
The three subgroups of Sco~OB2 are located at distances of 145,
140, and 118~pc. We assume an age of 5~Myr for US
\citep{degeus1989, preibisch1999}, 
and 20~Myr for UCL and LCC \citep{mamajek2002}.
Using the isochrones described in \S~\ref{sec: masses}, we find
that stars with $K_S=12~{\rm mag}$ correspond to stars with masses of 
0.08~$~{\rm M}_\odot$, 0.18~$~{\rm M}_\odot$, and
0.13~$~{\rm M}_\odot$ for US, UCL, and LCC, respectively.
Using the $K_S=12~{\rm mag}$ criterion 
we conclude that 77 out of 151 components are likely
background stars. The remaining 74 components are thus candidate companion stars.

Although this $K_S=12~{\rm mag}$ criterion works well for stars,
it is not very accurate below the hydrogen burning limit.
\cite{martin2004} found 28 candidate brown dwarf members of the US
subgroup using $I$, $J$, and $K$ photometry from the DENIS database. All
members have $10.9~{\rm mag} \leq K \leq 13.7~{\rm mag}$ 
and their spectral types are in the range
M5.5--M9. According to the $K_S=12~{\rm mag}$ criterion, 
19 of these brown dwarfs would be classified
as companion stars and 8 as background stars 
(one brown dwarf does not have a $K$ magnitude entry).
A significant fraction of possible brown dwarf companion stars
to our observed target stars
would be classified as background stars.

In this study we cannot determine with absolute certainty 
whether a detected component is a companion star or a 
background star. For example, a foreground star with 
$K_S > 12$~mag will not be excluded by the criterion described 
above. 
However, the differences between the companion 
star distribution and the background star distribution can be used to 
find out if the results are plausible or not.
For this purpose, we derive
the expected distribution of background stars per unit of angular 
separation $P(\rho)$, position angle $Q(\varphi)$,
and $K_S$ band magnitude $R(K_S)$ for the background stars.

Background stars are expected to be uniformly distributed over the image field with a surface density of $\Sigma$ stars per unit of solid angle.  The number of observed background stars $P(\rho)$ per unit of angular separation $\rho$ is given by
\begin{equation} \label{eq: theobackgroundgeneral}
P(\rho) = \frac{{\rm d}}{{\rm d}\rho} \int_{\rm \Omega_\rho} \Sigma \  d\Omega_\rho \ ,    
\end{equation}
where $\Omega_\rho$ is the part of the field of view contained within a circle with radius $\rho$ centered on the reference object (the target star). In calculating the expected $P(\rho)$ for our observations, we assume that the reference object is located in the center of the field of view. For a square field of view with dimension $L$, Equation~\ref{eq: theobackgroundgeneral} then becomes
\begin{equation} \label{eq: theobackground}
  P(\rho) \propto \left\{ 
  \begin{array}{llcccccc}	
    2 \pi \rho                               & {\rm for} & 0   & < & \rho &<& L/2&\\
    8\rho \left(\frac{\pi}{4} - \arccos \frac{L}{2\rho}\right)  
    & {\rm for} & L/2 & < & \rho &<& L/\!\sqrt{2}&.\\
    0                                        & {\rm for} &     &   & \rho &>& L/\!\sqrt{2}&\\
  \end{array} \right.
\end{equation}
For our observing strategy (Figure~\ref{figure: mosaic})
we have $L=\frac{3}{2} \cdot 12.76'' = 19.14''$.
For $\rho \leq L/2 = 9.6''$ the background star density for
constant $K_S$ is proportional to $\rho$, after which it decreases down to
$0$ at $\rho = L/\!\sqrt{2}  = 13.6''$. 
Here we have made the assumption that 
each target is exactly in the center of each quadrant.

We can also calculate $Q(\varphi)$, the expected 
distribution of background stars 
over position angle $\varphi$.
For $\rho \leq L/2$,  $Q(\varphi)$ is expected
to be random. For $L/2 < \rho \leq L/\!\sqrt{2}$ one expects
$Q(\varphi)=L^2/8 \cos^2\varphi$ for 
$0^\circ \leq \varphi < 45^\circ$~\footnote{This formula is 
also valid for $90^\circ \leq \varphi < 135^\circ$,
$180^\circ \leq \varphi < 125^\circ$, and $270^\circ \leq \varphi < 315^\circ$. 
$Q(\varphi)=L^2/8 \sin^2\varphi$ for all other position angles between $0^\circ$ 
and $360^\circ$.}. 
$Q(\varphi)$ is undefined for $\rho>L/\!\sqrt{2}$.
The above calculations apply to an ideal situation of our observing strategy. 
In practice, the target star is not always exactly centered in one detector 
quadrant, which influences $P(\rho)$ and $Q(\varphi)$ for $\rho \ga L/\!\sqrt{2}$. 

\cite{shatsky2002} analysed the background stars in their sky frames taken
next to their science targets in Sco~OB2. 
We use their data to derive $R(K_S)$, the expected
$K_S$ magnitude distribution for background stars.
We fit a second order polynomial to the
logarithm of the cumulative $K_S$ as a function of $K_S$ of the background
stars \citep[Figure~4 in][]{shatsky2002}. 
In our fit we only include background stars with 
$12~{\rm mag} < K_S < 16~{\rm mag}$ since
that is the $K_S$ magnitude range of the presumed background stars.
The $K_S$ magnitude distribution
$f(K_S)$ is then given by the derivative of the cumulative $K_S$
distribution. 

Figure~\ref{figure: separationhisto} shows the
companion star and background star distribution 
as a function of angular separation.
The two distributions are clearly different. The
companion stars are more centrally concentrated than the
background stars.
The solid curve represents $P(\rho)$, normalized in such a way that 
the number of stars corresponds to the expected number of background stars
with $12~{\rm mag}<K_S<16~{\rm mag}$
(see below and Figure~\ref{figure: expectedbackground}).
More background stars are present at angular separation $4''<\rho<6''$
relative to what is expected. This is confirmed with a
Kolmogorov-Smirnov (KS) test.
The KS test is usually 
quantified with the ``KS significance'', which ranges from 0 to 1 and corresponds to the 
probability that the two datasets are drawn from the same underlying distribution. 
For this KS comparison we only consider those background stars
with $\rho \leq L/\!\sqrt{2}= 9.6''$. For background stars with $\rho > 9.6''$
we expect the predicted distribution to differ slightly from 
the observed distribution (see above).
We find a KS significance level of $1.8 \times 10^{-4}$,
which implies that the excess of background stars with  
$4''<\rho<6''$ is real, assuming that the simple model
for the angular separation distribution is correct.
The excess of close background stars might be
caused by faint ($K_S > 12~{\rm mag}$) companion stars which
are misclassified as background stars.

The position angle distribution of companion stars
and background stars is shown in Figure~\ref{figure: compsfield}.
The objects are expected to be randomly distributed over
position angle for $\rho \leq L/2 = 9.6''$, 
which is illustrated in Figure~\ref{figure: compsfield}.
For objects with  $\rho \leq L/2 = 9.6''$ 
we find that, as expected,  the position angle distribution is random, 
for both the companion stars (KS significance level 0.86)
and the background stars (KS significance level 0.84). For
angular separations larger than $9.6''$ the objects are primarily found
in the 'corners' of the image 
($\varphi = 45^\circ, 135^\circ, 225^\circ, 315^\circ$).

The expected number of background stars per unit of angular
separation and unit of magnitude, $R(K_S)P(\rho)$ is plotted in Figure~\ref{figure:
expectedbackground}. We observe 19 background stars in the region $14~{\rm mag} \leq K_S
\leq 15~{\rm mag}$ and $5'' \leq \rho \leq 13''$ and normalize the theoretical
background star density accordingly. The brightness of the
PSF halo can explain the smaller number of background stars than expected for
$\rho \la 2.5''$ and $K_S \ga 15~{\rm mag}$. The predicted strong
decline in background stars for $\rho \ga 10''$ is present in
the observations. Few background objects with $12~{\rm mag}<K_S<14~{\rm mag}$ 
are expected in the angular separation range of $2''-4''$. 

\begin{figure}[bt]
  \includegraphics[width=0.5\textwidth,height=!]{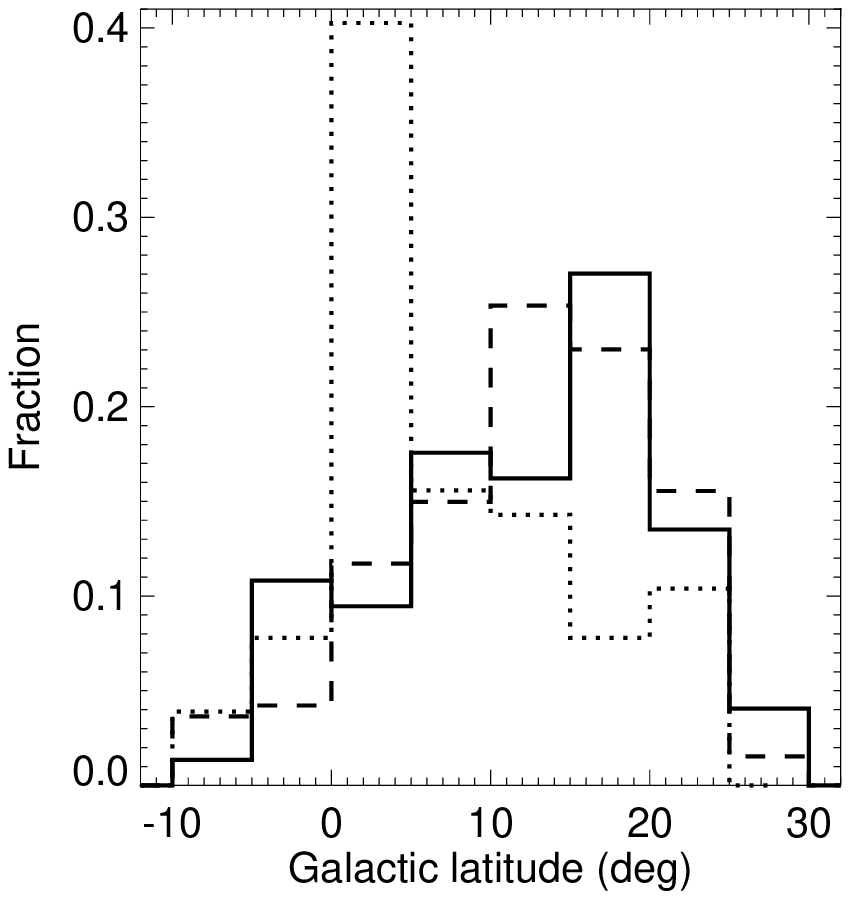} 
  \caption{The distribution of the 74 detected candidate companion stars (solid
  histogram) and 77 presumed background stars (dotted histogram) 
  as a function of Galactic latitude.
  The dashed histogram represents the distribution
  of {\it Hipparcos} member stars over Galactic latitude.
  The bin size is $5^\circ$ for the three histograms.
  The distributions are clearly different: the companion star and
  {\it Hipparcos} member star distribution peak at about the central Sco~OB2
  latitude. As expected, the background star distribution peaks at the 
  location of the Galactic plane.
  \label{figure: galacticplanehisto}}
\end{figure}

The distribution of the detected candidate companions and background stars on
the sky is shown in Figure~\ref{figure: galacticplane}. 
The covered fields close to
the Galactic plane include a larger number of background stars, 
as expected.
This is even better visible in Figure~\ref{figure: galacticplanehisto},
which shows the distribution of observed objects over Galactic latitude.
The distribution of Sco~OB2 {\it
Hipparcos} member stars over Galactic latitude is similar to that of the
companion stars.

In principle, it is also possible that another (undocumented) member star of
Sco~OB2 projects close to the target star. The status of these stars 
(companion star or ``background star'') cannot be determined with
the CMD or the $K_S=12$~mag criterion.
\cite{preibisch2002} estimate the
number of member stars in the US subgroup in the range $0.1-20~{\rm M}_\odot$ to be
about 2500. This corresponds roughly to a stellar surface density of $8\times
10^{-7}$ star\,arcsec$^{-2}$. The chance of finding another US member star
(i.e. not a binary companion) in a randomly pointed observation of US is of
the order $0.03\%$. The detection probabilities for the other two subgroups of
Sco~OB2 are similarly small and negligible for our purpose.

\section{Properties of the companion stars} \label{sec: properties}

\subsection{General properties} \label{sec: generalproperties}

The angular separation distribution of the companion stars is 
centrally concentrated (Figures~\ref{figure: separationhisto} 
and~\ref{figure: compsfield}). 
The error in
the angular separation is typically $0.0015''$. 
Faint ($K_S \ga 11~{\rm mag}$) objects with angular separation less than $\rho \approx
0.8''$ cannot be detected in the presence of the PSF halo of the bright
primary star, while the upper limit of $13.5''$ is determined by
the field of view. The distribution over position
angle is random for $\rho \leq L/2 = 9.6''$, as expected 
(see \S~\ref{sec: backgroundstars}). 
The median error in position angle is
$0.03^\circ$, increasing with decreasing angular separation.

\begin{figure}[bt]
  \begin{tabular}{cc}
    \includegraphics[width=0.22\textwidth,height=!]{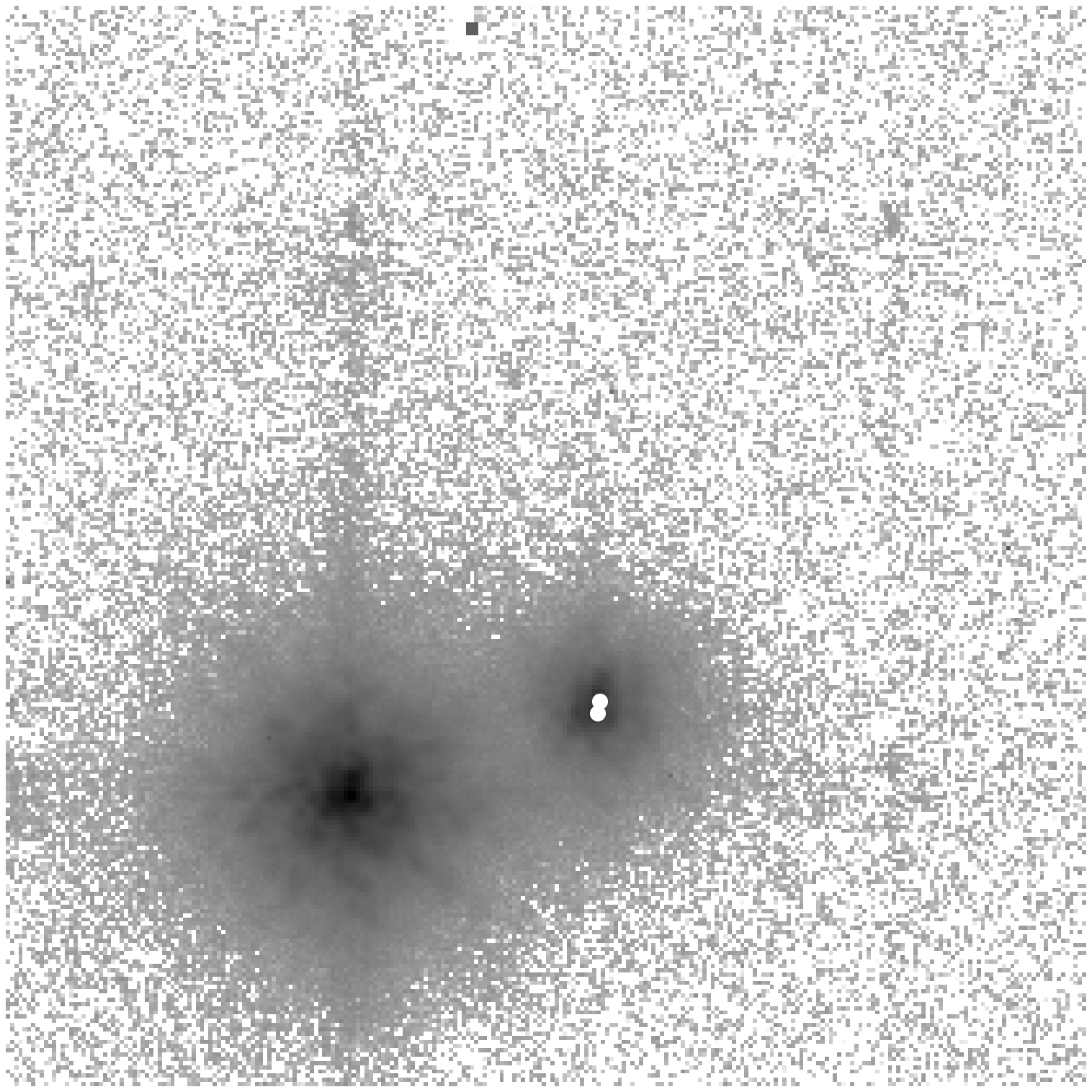} &
    \includegraphics[width=0.22\textwidth,height=!]{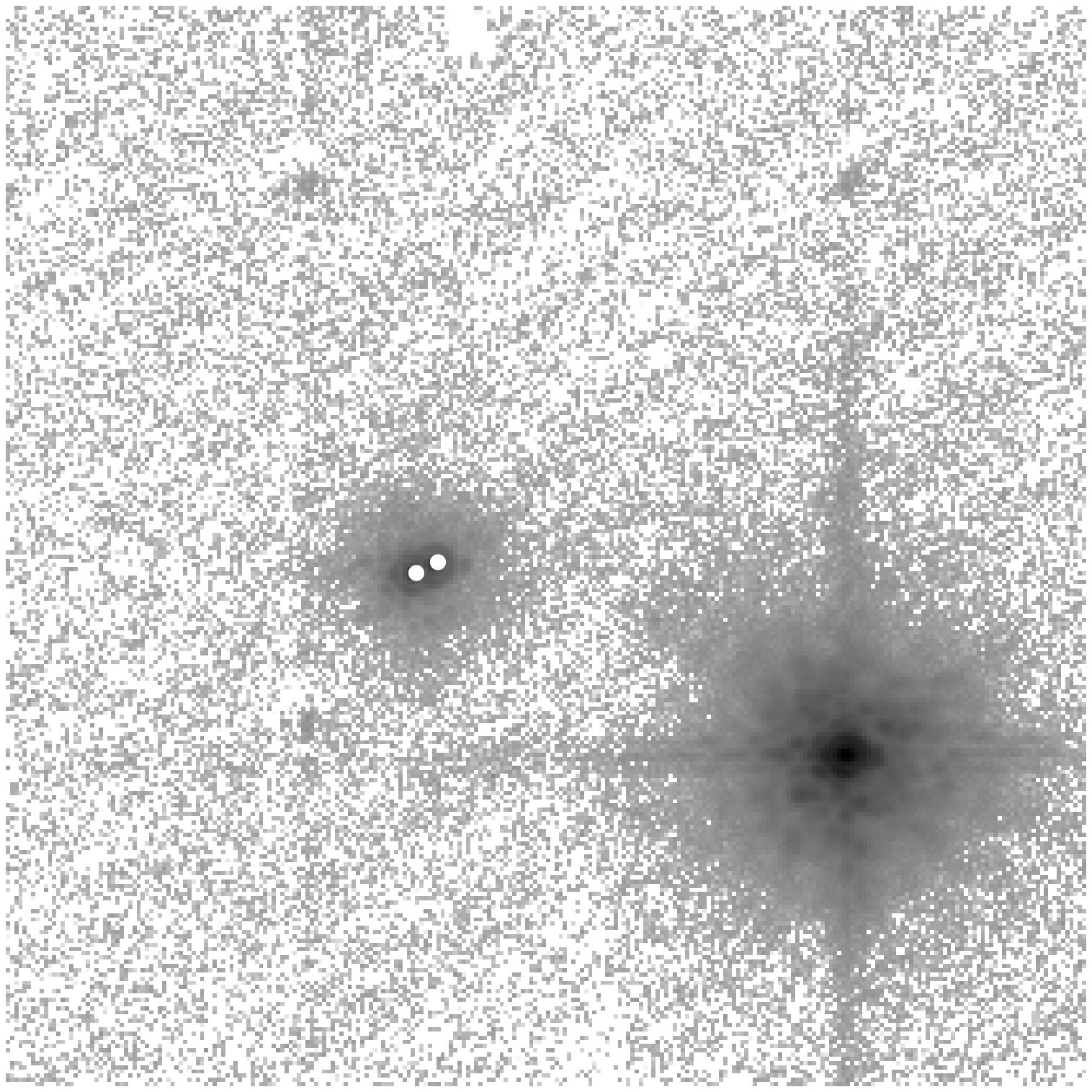} \\
  \end{tabular}
  \caption{
    {\it Left:} The hierarchical triple HIP68532. Both companion stars
    (indicated with the white dots) were previously undocumented.
    {\it Right:} HIP69113, another multiple system, where
    both close companion stars (white dots) 
    were previously undocumented.
    A third known companion star with angular separation of 
    $28''$ is not detected in our survey because of
    the large angular separation.
    \label{figure: triples}}     
\end{figure}

Two observed systems are clearly hierarchical triples where
a double 'secondary' orbits the primary star. These are HIP68532, where
the double 'secondary' has  $\rho=3.1'' $ and $\varphi=290^\circ$, and 
HIP69113, with $\varphi=65^\circ$ and $\rho=5.4''$.
HIP69113 has a third companion star 
with $\rho=28.6'' $ and $\varphi=35^\circ$ \cite{duflot1995}.
This star is not observed in our survey because of the large distance
from the primary star. 
The nature of the HIP69113 system is extensively discussed
by \cite{huelamo2001}.
Figure~\ref{figure: triples} shows
HIP68532 and HIP69113 with companion stars.
The target stars HIP52357, HIP61796, and HIP81972 also
have two companion stars. The two companion stars of each of
these systems have approximately the same angular separation and
position angle, but it is unclear
whether these systems are hierarchical in the way described above.

The 199 observed primaries have a $K_S$ magnitude range of $5.2-8.3~{\rm mag}$. The $K_S$
magnitudes of the companion stars (Figure~\ref{figure: detectionlimits}) range
from $K_S = 6.4~{\rm mag}$ to $12.0~{\rm mag}$, 
the upper limit resulting from the criterion that
was used to remove the background stars from the sample. The
primaries in our sample all have similar spectral type, distance, and
interstellar extinction. The magnitude of the companions relative to that of their
primaries, $\Delta K_S$, spans the range from $0.0~{\rm mag}$ to $6.0~{\rm mag}$. 
The median error is $0.05~{\rm mag}$ in $J$, $0.04~{\rm mag}$ in $H$,
$0.07~{\rm mag}$ in $K_S$, and $0.075~{\rm mag}$ in $J-K$.

Properties of the 199 target stars, 74 candidate companion stars, and 77
presumed background stars are listed in Table~\ref{table: results}. The name of
the star is followed by the $J$, $H$, and $K_S$ magnitude. If the magnitude
is derived from measurements done under non-photometric conditions, a remark
is placed in the last column. The spectral type of each primary is taken from
the \textit{Hipparcos} catalogue. The angular separation and position angle
(measured from North to East) are derived from the combination of all
available observations for a particular star. For each star we also list the
status (p = primary star, 
c = candidate companion star, nc = new candidate
companion star, b = background star) and the subgroup membership.

\begin{table*}[bth] 
\begin{centering}
\begin{tabular}{|cc cc cc cc cc|}
  \hline
  Star & $J$ mag & $H$ mag & $K_S$ mag & spectral type & $\rho$ ($''$)  & $PA$ ($^\circ$) & status & subgroup & remarks\\ 
  \hline 
  \hline 
HIP50520 &  &  & 6.23 & A1V &  &  & p & LCC &  \\
 &  &  & 6.39 &  & 2.51 & 313.3 & c &  &  \\
\hline
HIP52357 &  &  & 7.64 & A3IV &  &  & p & LCC &  \\
 &  &  & 7.65 &  & 0.53 & 73.0 & c &  &  \\
 &  &  & 11.45 &  & 10.04 & 72.7 & nc &  &  \\
\hline
HIP53524 &  &  & 6.76 & A8III &  &  & p & LCC &  \\
 &  &  & 12.67 &  & 4.87 & 316.9 & b &  &  \\
\hline
HIP53701 & 6.30 & 6.37 & 6.48 & B8IV &  &  & p & LCC &  \\
 & 9.05 & 8.76 & 8.86 &  & 3.88 & 75.8 & c &  &  \\
 & 13.06 & 12.93 & 13.04 &  & 6.57 & 120.1 & b &  &  \\
\hline
HIP54231 &  &  & 6.75 & A0V &  &  & p & LCC &  \\
\hline
HIP55188 &  &  & 7.43 & A2V &  &  & p & LCC &  \\
\hline
HIP55899 &  &  & 7.07 & A0V &  &  & p & LCC &  \\
\hline
HIP56354 &  &  & 5.78 & A9V &  &  & p & LCC &  \\
  \hline
  \dots & \dots & \dots &\dots  & \dots & \dots & \dots & \dots & \dots & \dots \\
  \hline
  \end{tabular}
  \caption{Results from the ADONIS adaptive optics survey among 199 Sco~OB2
    member stars. The columns list: {\it Hipparcos} number, observed $J$, $H$ and $K_S$
    magnitude, spectral type of the primary, angular separation (arcsec),
    position angle (degrees; measured from North to East), and status (p = primary; c =
    companion star; nc = new companion star; b = background star). The last two
    columns list the subgroup of which the system is a member (US = Upper
    Scorpius; UCL = Upper Centaurus Lupus; LCC = Lower Centaurus Crux), and 
    a flag in case the observing conditions were not
    photometric. The {\it Hipparcos} member stars HIP77315 and HIP77317
    are a common proper motion pair (see the Double Star Catalog).
    In this paper the two stars are treated as separate primaries and not 
    as one system.
    Table 1 is presented in its entirety in the electronic edition of
    Astronomy \& Astrophysics. A portion is shown here for guidance regarding its form
    and content. \label{table: results}}
\end{centering}
\end{table*}

\subsection{Color-magnitude diagram and isochrones} \label{sec: hrdiagram}

\begin{figure*}[bt]
  \centering
  \includegraphics[width=\textwidth,height=!]{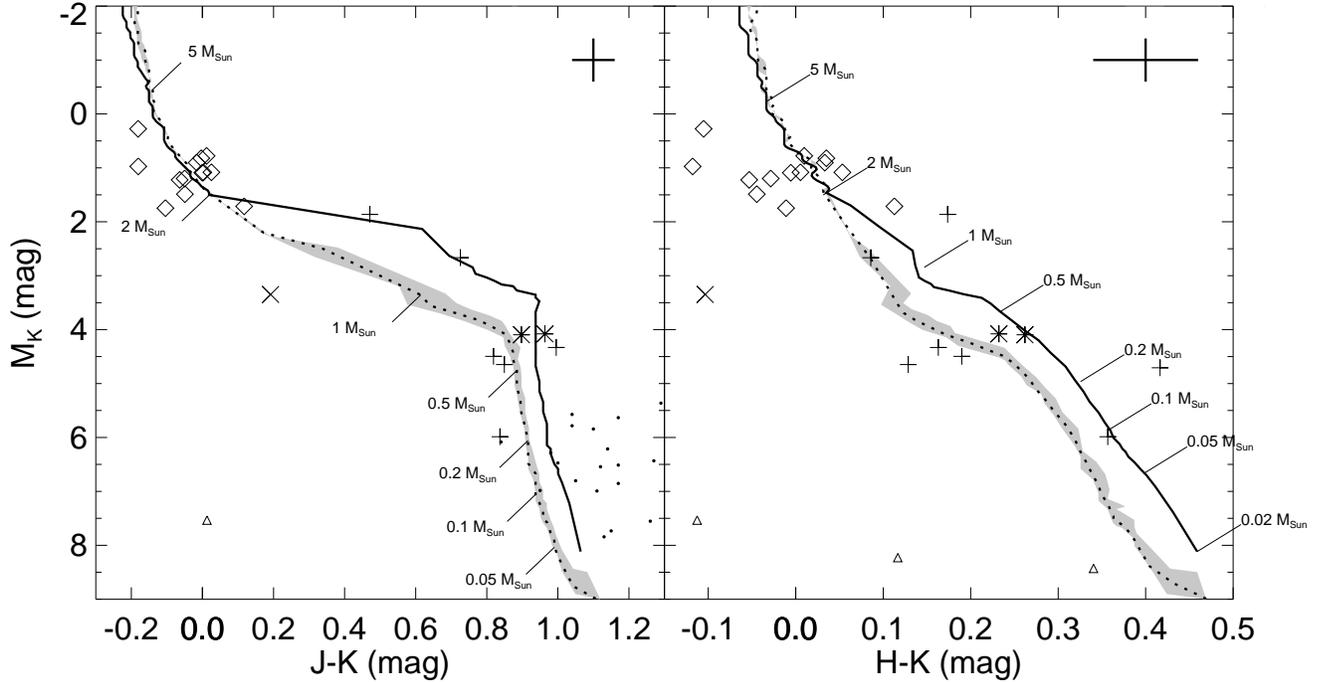}
  \caption{The color-magnitude diagram of the primaries,
    companion stars and background stars for which we obtained
    ADONIS/SHARPII+ multi-color AO observations.
    The $M_{K_S}$ magnitudes are corrected for distance and
    extinction. Companion stars in the three different subgroups are indicated
    with plusses (US), asterisks (UCL) and crosses (LCC). Target stars and
    background stars are indicated with diamonds and triangles, respectively
    (see \S~\ref{sec: backgroundstars}). The curves represent isochrones of
    5~Myr (solid curve) and 20~Myr (dotted curve). 
    The 15~Myr and 23~Myr isochrones enclose the gray-shaded area
    and represent the uncertainty in the age of the
    UCL and LCC subgroups.
    The mass scale is indicated for the 20~Myr isochrone ({\it left})
    and the 5~Myr isochrone ({\it right}).
    The brown dwarf candidates
    identified by \cite{martin2004} are indicated as dots. 
    The median formal errors are indicated 
    as error bars in the top-right corner of each plot.
    The locations of the target stars and companion stars are 
    consistent with the isochrones (within the error bars).
    The presumed background stars are located far away from the 
    isochrones, implying that our criterion to separate
    companion stars and background stars is accurate.
    \label{figure: 3hr}}
\end{figure*}

\begin{figure}[bt]
    \includegraphics[width=0.5\textwidth,height=!]{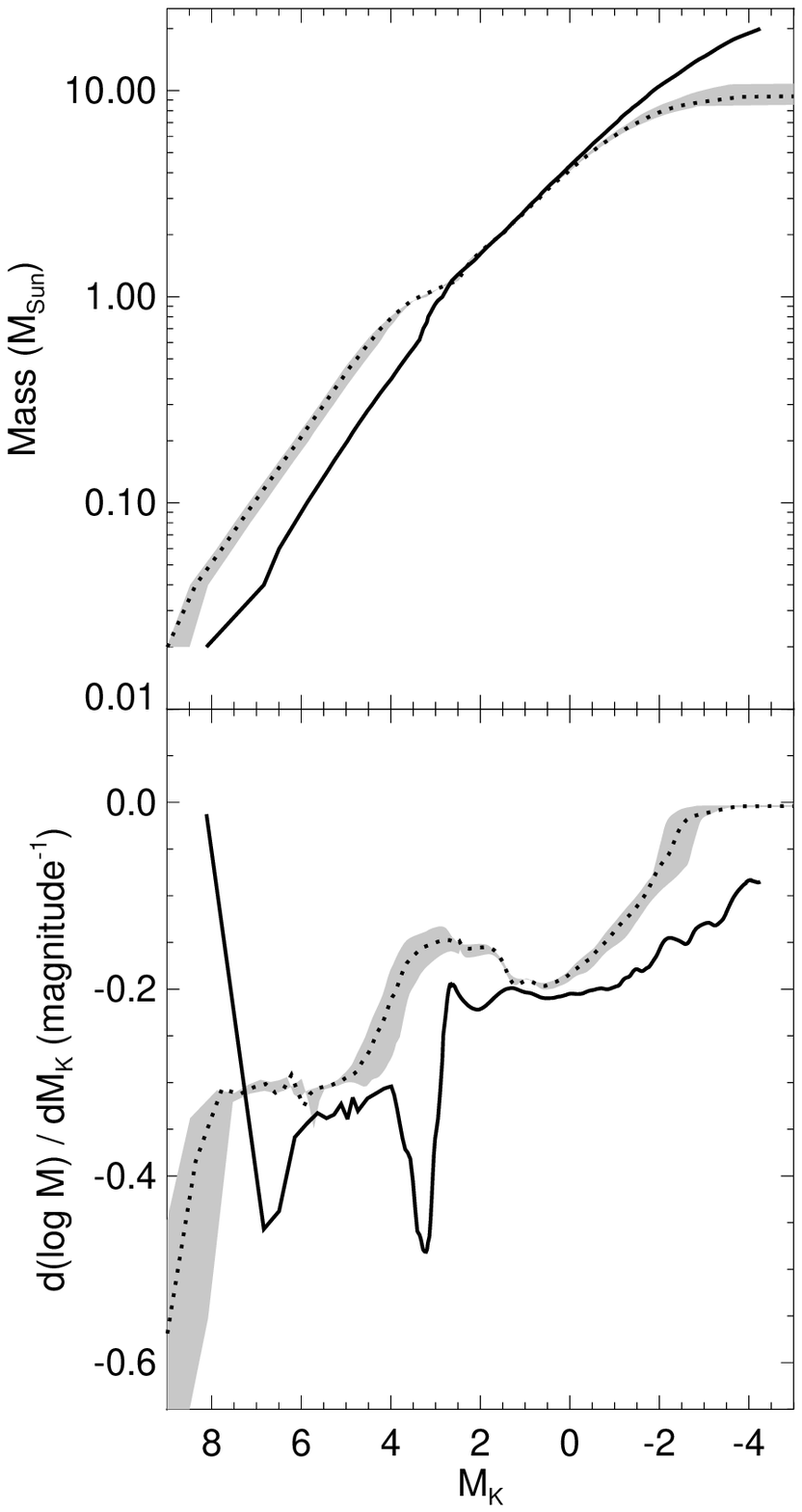}
    \caption{{\it Top:} The relation between mass and $M_{K_S}$ magnitude. The
    5~Myr isochrone (solid curve) 
    is used to find masses of the companion
    stars in US, while the 20~Myr isochrone (dotted curve) 
    is used for the UCL and LCC subgroups. 
    The gray-shaded area is enclosed by the 15~Myr and 23~Myr
    isochrones, and represents the uncertainty in the age of the
    UCL and LCC subgroups.
    For objects at the mean Sco~OB2 distance of 130~pc,
    $M_{K_S}=6$ corresponds to $K_S=11.6$. {\it Bottom:} The derivative of the
    mass-magnitude relation. The conversion from $M_{K_S}$ to mass is most
    accurate where the absolute value of the derivative 
    ${\rm d}M/ {\rm d}M_{K_S}$ is small. \label{figure:
    massmagnituderelation}}
\end{figure}

In \S~\ref{sec: backgroundstars} we mentioned that it is
possible to separate background stars and companion stars
using their position in the CMD. In this section we 
derive the absolute magnitude $M_{K_S}$ for all observed objects.
We additionally derive colors for objects with multi-color 
observations. Only measurements obtained under photometric 
conditions are used. 
We construct a CMD  and
determine whether the observed components are background stars
or companion stars
by comparing their position in the CMD to that of the isochrones.
In \S~\ref{sec: masses} we will 
use the absolute magnitude $M_{K_S}$ and the
isochrones to derive the masses of the primary and
companion stars.

We calculate the absolute magnitude $M_{K_S}$ 
using the distance $D$ and extinction
$A_{K_S}$ for each target star individually. 
In \cite{debruijne1999} the distance to the Sco~OB2
member stars is derived from the 
secular parallax $\pi_{\rm sec}$. 
The secular parallax is calculated from the observed positions and
proper motions of member stars, where the fact that 
all stars in an association 
share the same space motion is exploited. The secular parallax 
can be up to more than two times as precise compared to the
{\it Hipparcos} parallax. 
Secular parallaxes with $g=9$ are used when
available, and with $g=\infty$ otherwise 
\citep[$g$ measures the model-observation discrepancy; see][for
details]{debruijne1999}. The interstellar extinction is also taken from
\cite{debruijne1999}. For the stars that do not have an extinction entry in
\cite{debruijne1999}, an estimate of the extinction is calculated using:
\begin{equation}
A_{K_S} = R_V \times E(B-V) / 9.3,
\end{equation}
where $R_V=3.2$ is the standard ratio of total-to-selective extinction
\citep[see, e.g.,][]{savage1979}. The value 9.3 is the ratio between $V$ band
and $K_S$ band extinction \citep{mathis1990}, and $E(B-V) = (B-V)-(B-V)_0$ is
the color excess. $(B-V)$ is the color listed in \cite{debruijne1999}. The
theoretical color $(B-V)_0$ is derived from the spectral type using the
broadband data for main sequence stars in \cite{kenyon1995}. 
The extinction for the UCL member HIP68958
(spectral type Ap...) is not listed in \cite{debruijne1999} and could not be
derived from \cite{kenyon1995}. We therefore take 
$A_{V,{\rm HIP68958}} = 0.063~{\rm mag}$. This is
the median $A_V$ value for those UCL member
stars which have full photometric data in \cite{debruijne1999}.

Only for a subset of observed targets observations in
more than one filter are available; 
the corresponding CMDs are plotted in Figure~\ref{figure: 3hr}. 
The observed primary stars are all of similar spectral type but show 
scatter in $M_{K_S}$, $J-K$, and $H-K$.  This
can be explained by the errors in near-infrared photometry, which are
typically $0.07~{\rm mag}$ for our observations. 
The uncertainties in the parallax
and reddening introduce an additional error when 
calculating the absolute magnitudes. Median
errors in $M_J$ and $M_K$ are consequently about $0.36~{\rm mag}$. The colors
$J-K$ and $H-K$ have median errors of $0.02~{\rm mag}$ and $0.06~{\rm mag}$,
respectively.

Given the age of the three Sco~OB2 subgroups, 
not all objects are expected to be positioned on the
main sequence. For the young age of Sco~OB2, stars of spectral type~G or later
have not reached the main sequence yet, while the O~stars have already
evolved away from the main sequence.
The age differences between the Sco~OB2 subgroups are relatively
well-determined. Different values for the absolute ages are
derived from the kinematics of the
subgroups \citep{blaauw1964,blaauw1978} and stellar evolution
\citep{dezeeuw1985,degeus1989,preibisch2002,mamajek2002,sartori2003}. 
The age of the US subgroup is 5~Myr, without a
significant age spread 
\citep{dezeeuw1985,degeus1989,preibisch2002}.
\cite{mamajek2002} derive an age between 
15~and 22~Myr for members of UCL
and 17~and 23~Myr for members of LCC. 
In this paper we adopt an age 
of 20~Myr for both UCL and LCC.

We construct isochrones which are very similar to
those described in \cite{preibisch2002}.
For the low-mass stars ($M < 0.7\, {\rm M}_\odot$) 
we use exactly the same isochrone as \cite{preibisch2002}:
we use the models from \cite{chabrier2000}, which are
based on those from \cite{baraffe1998} 
For $0.02 \leq M/{\rm M}_\odot < 0.7$ we take the models 
with $[\rm{M/H}]=0$ and $L_{\rm mix} = H_P$, 
where $L_{mix}$ is the mixing length and $H_P$ the pressure scale height.
For $0.7 \leq M/{\rm M}_\odot < 1$ we use the models with
$[\rm{M/H}]=0$ and $L_{\rm mix} = 1.9 H_P$.
We use the models described in \cite{palla1999} for 
$1 \leq M/{\rm M}_\odot < 2$. Near-infrared magnitude tracks,
derived using the procedure described in \cite{testi1998},
were kindly provided by F.~Palla.
Finally, for $M/{\rm M}_\odot > 2$ we use the models from 
\cite{girardi2002} \citep[based on][]{bertelli1994} 
with $Y = 0.352$ and $Z = 0.05$, where
$Y$ and $Z$ are the helium and metal abundance, respectively.

The resulting 5~Myr isochrone (for US) 
and 20~Myr isochrone (for UCL and LCC) 
are plotted in Figure~\ref{figure: 3hr}. 
Both isochrones are continuous.
Figure~\ref{figure: massmagnituderelation} shows
the relationship between stellar mass and $K$-band
absolute magnitude for 5~Myr and 20~Myr
populations.
The gray-shaded area in Figures~\ref{figure: 3hr} 
and~\ref{figure: massmagnituderelation} is enclosed
by the 15~Myr and 23~Myr isochrones, and shows the 
effect of the age uncertainty in the UCL and LCC
subgroups. 

The positions of the primary and companion stars in the CMD 
are consistent with the
isochrones (except the companion of the LCC member star HIP53701). 
No obvious correlation of the position of these primaries in the
CMD with spectral type is present due to the errors in the colors and magnitudes.
The presumed background stars are located far from 
the isochrones in the CMD, 
supporting our hypothesis that they are indeed background stars.

In Figure~\ref{figure: 3hr} we added the 
candidate brown dwarf members in US from \cite{martin2004}.
We estimate the absolute magnitude of these brown dwarfs
using the mean US distance of 145~pc and the mean
extinction $A_K = 0.050~{\rm mag}$ (see below).
The colors and magnitudes of these objects are consistent with
the 5~Myr isochrone (solid curve), as is expected for US member stars.

\subsection{Masses and mass ratios} \label{sec: masses}

\begin{figure}[bt]
    \includegraphics[width=0.5\textwidth,height=!]{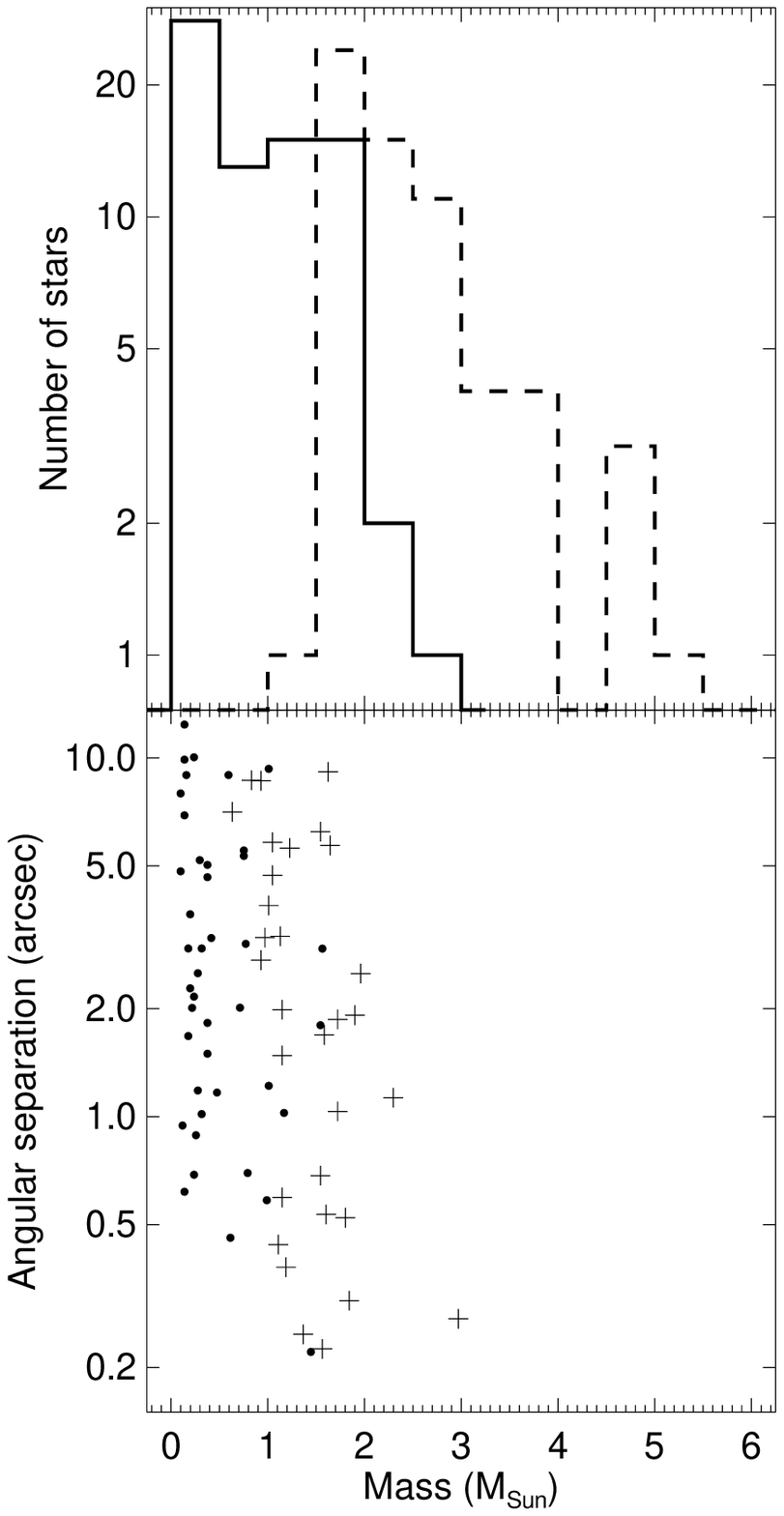}
    \caption{{\it Top:} The mass distribution of 
      the 74 companion stars (solid
      histogram) and corresponding primaries (dashed histogram)
      observed in our AO survey.
      The bin size is $0.5~{\rm M}_\odot$.
      Masses are derived from their $K_S$ magnitude 
      and age (see \S~\ref{sec: masses}). 
      The median errors in the mass are $0.4~{\rm M}_\odot$ for the primaries
      and $0.1~{\rm M}_\odot$ for the companions. 
      {\it Bottom:} Companion star mass versus angular separation. 
      New and previously known companion stars are indicated as
      dots and plusses, respectively.
      Low-mass close
      companion stars are not detected due to the PSF wings of the corresponding
      primary star. \label{figure: massdistributions} }
\end{figure}

We derive the mass of the primary and companion stars 
from their $M_{K_S}$ magnitude
using the isochrones described in \S~\ref{sec: hrdiagram}
(Figures~\ref{figure: 3hr} and~\ref{figure: massmagnituderelation}). 
We use the 5~Myr
isochrone for US and the 20~Myr isochrone for UCL and LCC. The conversion from
$M_{K_S}$ to mass is most accurate for B- and A-type stars,
where the absolute value of the derivative 
${\rm d}M/ {\rm d}M_{K_S}$ is small. 
We find masses between $1.4~{\rm M}_\odot$ and $7.7~{\rm M}_\odot$ for the primaries and masses
between $0.1~{\rm M}_\odot$ and $3.0~{\rm M}_\odot$ for the companions. 
Given the error $\Delta M_{K_S}$ on $M_{K_S}$, we calculate
for each star the masses $M_-$ and $M_+$ 
that correspond to  $M_{K_S}-\Delta M_{K_S}$
and $M_{K_S}+\Delta M_{K_S}$. We define the error on the mass
to be $\frac{1}{2} (M_+-M_-)$.
The median of the error in the mass, which is a lower limit for the
real error, is $0.4~{\rm M}_\odot$ for the primaries 
and $0.1~{\rm M}_\odot$ for the companions. 

Figure~\ref{figure: massdistributions} (top) shows the mass distribution of
the companion stars and of the primary stars 
to which they belong. The mass distribution
of the observed target stars without companions is similar to the latter.
The plot showing companion star mass as
a function of angular separation (Figure~\ref{figure: massdistributions},
bottom) closely
resembles Figure~\ref{figure: detectionlimits}.
The reason for this is that the companion star mass is
derived from the $K_S$ magnitude.

\begin{figure}[bt]
    \includegraphics[width=0.5\textwidth,height=!]{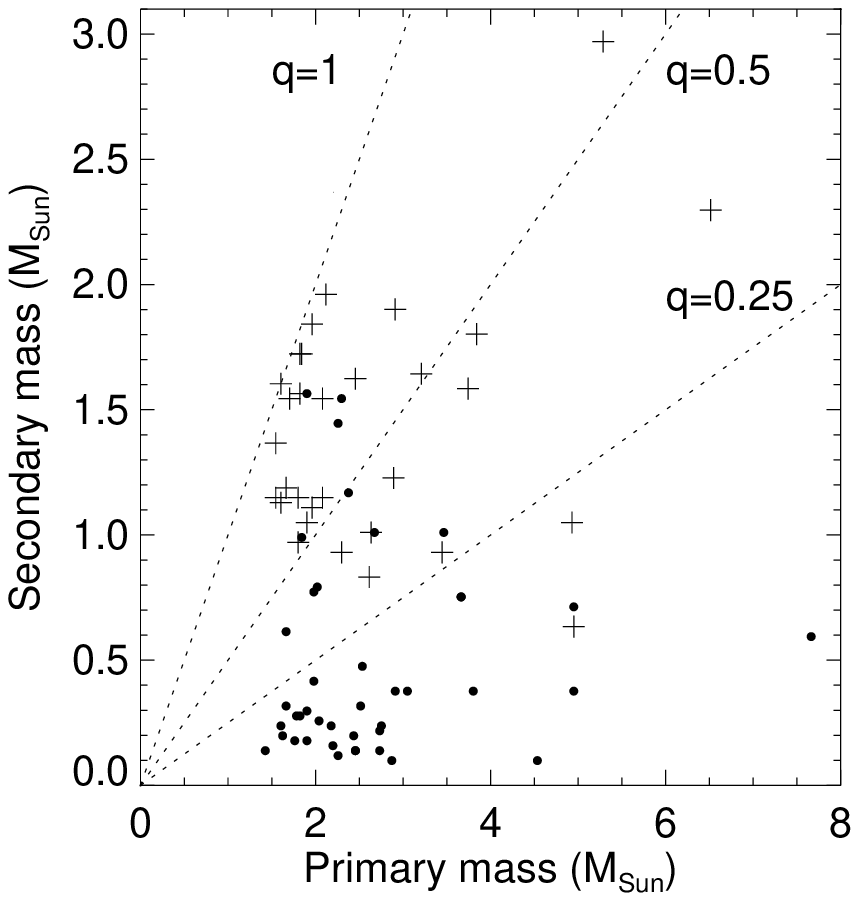} 
    \caption{Companion star mass versus primary star mass. The dashed lines
    represents the $q=0.25$, $q=0.5$, and $q=1$ binaries
    and are shown to guide the eye. 
    The 41 new companion stars
    and the 33 previously known companion stars are indicated with
    dots and plusses, respectively.
    The observed companion stars have masses lower than their associated
    primaries. Most systems with 
    previously undocumented companion stars
    have $q < 0.25$.
    \label{figure: masses}}
\end{figure}

The mass ratio $q$ and error are calculated for each primary-companion pair. 
The median formal error on the mass ratio is 0.07 and is to first order 
proportional to that of the companion star mass. The real error on the mass 
ratio is most likely larger than the formal error stated above.
As indicated by \cite{shatsky2002}, the mass ratio of a system primarily 
depends on the magnitude difference between primary and secondary. 

\begin{figure}[!hbt]
  \includegraphics[width=0.5\textwidth,height=!]{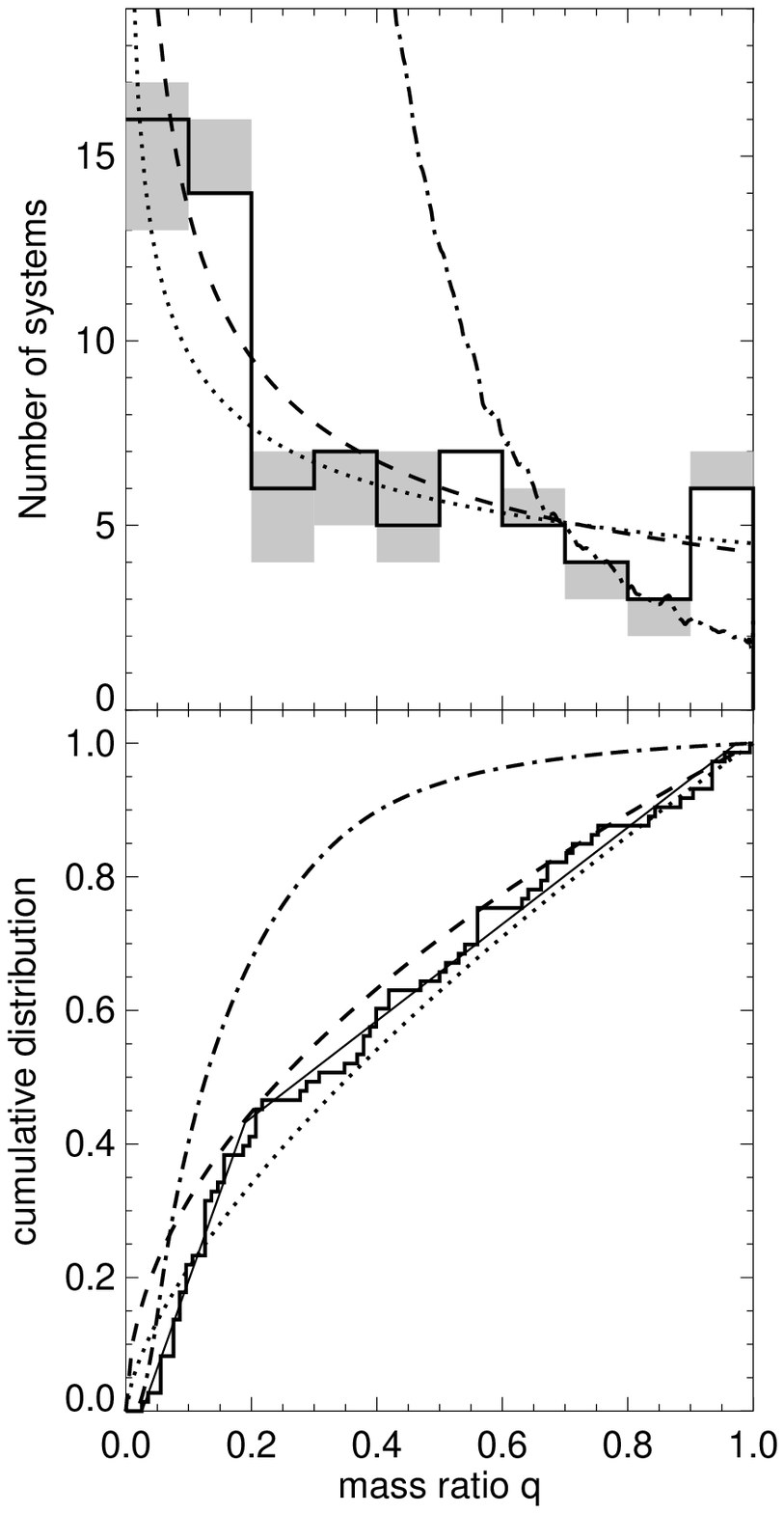}\hspace{0cm}
  \caption{{\it Top:} The mass ratio distribution for the 
    74 systems for which we observe companion stars
    in our near-infrared AO survey (histogram).
    The mass ratio is defined as $q = M_c/M_p$ where
    $M_c$ is the companion star mass and $M_p$ the mass of the 
    corresponding primary star. All primaries are 
    {\it Hipparcos} member stars with A and late-B spectral type.
    The $q$ distribution does not change significantly
    if ages of 15~Myr or 23~Myr for UCL and LCC are assumed 
    (gray-shaded areas).
    The mass ratio distributions $f(q) = q^{-\Gamma}$ are represented by the
    curves. 
    For our observations we find $\Gamma = 0.33$ (dotted
    curve). \cite{shatsky2002} find $\Gamma = 0.5$ (dashed curve). 
    The distribution which follows from random pairing
    (dash-dotted curve) is clearly excluded. 
    The curves are normalized using the observed
    mass ratio distribution for the 
    companion stars in the range $0.5 < q < 1.0$.
    {\it Bottom:} The observed cumulative mass
    ratio distribution (histogram). 
    The dotted and
    dashed curve are cumulative distribution functions corresponding to
    $\Gamma = 0.33$ and $\Gamma = 0.5$, respectively. 
    The fitted function described in Equation~\ref{eq: twolinefit} is 
    shown as the thin solid curve. The dash-dotted
    curve represents the cumulative $q$ distribution for random
    pairing. 
    \label{figure: massratiohisto}}
\end{figure}

The mass ratio distribution is shown in Figures~\ref{figure: masses} and~\ref{figure:
massratiohisto}. The smallest mass ratios are found for the companions of
HIP80474 ($q=0.022 \pm 0.008$) and HIP77911 ($q=0.035 \pm 0.012$). 
This corresponds well with the
expected minimum value for $q$: a typical primary in our sample has spectral
type A0~V ($2~{\rm M}_\odot$), while the least massive companions considered have
$K_S=12~{\rm mag}$ ($\sim 0.1~{\rm M}_\odot$), giving $q=0.05$. 
The systems with large mass ratio 
($q>0.9$) are HIP80324 ($q=0.91 \pm 0.23$), HIP50520
($q=0.93 \pm 0.19$), HIP80238 ($q=0.94 \pm 0.29$), HIP64515 ($q=0.94 \pm 0.15$), 
HIP61639 ($q=0.95 \pm 0.15$), and HIP52357 ($q=1.00 \pm 0.34$). 
Note that for these systems the possibility that $q>1$ is included. 
Due to the error in the mass it is not possible to say which star 
in these systems is more massive, and hence, which is the primary star.
Although the uncertainty in the age of UCL and LCC is not negligible,
the effect on the mass ratio distribution is small 
(gray-shaded areas in Figure~\ref{figure: massratiohisto}).

Following \cite{shatsky2002} we fit a power-law of the form $f(q) = q^{-\Gamma}$
to the mass ratio distribution. We find that models
with $\Gamma = 0.33$ fit our observations best (KS significance
level 0.33). 
The error in $\Gamma$ due to the age uncertainties in 
the UCL and LCC subgroups is 0.02.
This is in good agreement with \cite{shatsky2002},
who surveyed 115 B-type stars in Sco~OB2 for binarity
and observed 37 physical companions around these stars. 
They find $\Gamma = 0.3 - 0.5$ for their mass ratio distribution.
We observe an excess of systems with $q \sim 0.1$ with
respect to what is expected for models with $\Gamma=0.33$.
This could partially be explained by bright background stars that are
misclassified as companion stars, but this is unlikely to be an important
effect (see \S~\ref{sec: backgroundstars}). 
The observational biases (e.g. the detection limit) are not 
taken into account here. For a more detailed description of the
effect of observational biases
on the mass ratio distribution 
we refer to \cite{hogeveen1990} and \cite{tout1991}.

The cumulative mass ratio distribution $F(q)$ can also
be described as a curve consisting of two line segments.
We fit and find that the 
mass ratio distribution
closely follows the function
\begin{equation} \label{eq: twolinefit}
  F(q) = \left\{
  \begin{array}{lll}
    2.63 q - 0.067 & {\rm \quad for \quad } & 0.03 < q \leq 0.19 \\
    0.72 q + 0.294 & {\rm \quad for \quad } & 0.19 < q \leq 0.97, \\
  \end{array}
  \right.
\end{equation}
with a root-mean-square residual of $0.019$ 
(see Figure~\ref{figure: massratiohisto}).

We investigate whether the observed mass ratio distribution 
could be the result of random pairing between primary stars
and companion stars, such as observed for solar-type stars
in the solar neighbourhood \citep{duquennoy1991}.
To this end, we use Monte Carlo simulations to calculate the mass ratio distribution
that is expected for random pairing. 

The current knowledge about the brown dwarf
population in Sco~OB2 is incomplete \citep[e.g., Table 2 in][]{preibisch2003}.
Therefore we make a comparison with the 
mass ratio distribution resulting from two different 
initial mass functions (IMFs)
(IMF$_{-0.3}$ and IMF$_{2.5}$),
which differ in slope ${\rm d}N/{\rm d}M$
for substellar masses.
\begin{equation} \label{equation: imf1}
  {\rm IMF_\alpha:~~} \frac{{\rm d}N}{{\rm d}M} \propto \left\{
  \begin{array}{llll}
    M^{\alpha}  & {\rm for \quad } 0.02 & \leq M/{\rm M}_\odot & < 0.08 \\
    M^{-0.9}  & {\rm for \quad } 0.08 & \leq M/{\rm M}_\odot & < 0.6 \\
    M^{-2.8}  & {\rm for \quad } 0.6  & \leq M/{\rm M}_\odot & < 2   \\
    M^{-2.6}  & {\rm for \quad } 2    & \leq M/{\rm M}_\odot & < 20, \\
  \end{array}
  \right.
\end{equation}
For $M \geq 0.10~{\rm M}_\odot$ IMF$_\alpha$ is equal to the IMF in US 
that was derived by \cite{preibisch2002}. The \cite{preibisch2002}
IMF is extrapolated down to $M = 0.08~{\rm M}_\odot$.
For $0.02 \leq M/{\rm M}_\odot < 0.08$ we consider 
$\alpha=-0.3$ \citep{kroupa2002} and $\alpha=2.5$ 
\citep[][Fit~I]{preibisch2003}. Most other models described in
\cite{preibisch2003} have $-0.3<\alpha<2.5$ 
for $0.02 \leq M/{\rm M}_\odot < 0.08$.

We draw the primary and secondary stars from
IMF$_\alpha$. 
Primary and secondary masses are independently drawn from the mass 
range $[0.02,20]~{\rm M}_\odot$ and are swapped
if necessary, so that the primary star is the most massive. Only
those systems with primary mass larger than  $1.4~{\rm M}_\odot$ 
and smaller than $7.7~{\rm M}_\odot$ are considered.
We calculate the resulting mass ratio distribution.
A KS comparison between the mass ratio distribution
resulting from random pairing and the observed distribution
shows that the random-pairing hypothesis can be rejected
for both IMF$_{-0.3}$ and IMF$_{2.5}$.
This is clearly visible in Figure~\ref{figure: massratiohisto}.

Random pairing requires more systems with low mass ratio than we observe.
The error in the mass ratio (ranging between 0.008 and 0.34) 
cannot compensate for the lack of faint companions. 
Even if all close ($\rho \leq 4$) background stars are treated as
low-mass companions, random pairing is excluded. 
Hence, there is a deficit of low-mass companions around
late-B and A type stars compared to what is expected from random pairing. 

Several observed primary stars are actually unresolved binary systems.
Since we derive the mass for this 'primary' using the total $K_S$ flux of such
an unresolved system, we introduce an error in the mass and hence the mass ratio.
Out of the 199 observed primary stars 18 are known 
spectroscopic binaries (14 in US, 3 in UCL, and 1 in LCC). 
These binary systems are too tight to be resolved in our AO survey.
We observe companion stars around three of these primaries:
HIP77911, HIP77939, and HIP79739 (all in US). For these systems
we expect the mass ratio, which we define here as the mass of the
companion star divided by the total mass of the inner part
of the system, to deviate significanly from the value 
derived from our near-infrared observations.


A closer inspection of Figure~\ref{figure: detectionlimits} reveals a lack of
stellar components with $\rho \leq 3.75''$ and $12~{\rm mag} \la K_S \la 14~{\rm mag}$. 
If companion stars or background stars
with these properties exist, we should have seen them. 
In \S~\ref{sec: backgroundstars} we showed that few background stars are 
expected in this region. For
example, the brown dwarfs (M5.5--M9) detected by \cite{martin2004} in US have
$11~{\rm mag} \la K \la 14~{\rm mag}$. If similar objects within $\rho \leq 3.75''$ were
present, they would have been detected.
This `gap' is not the result of observational biases: the detection limit shows
that objects with these properties are detectable. 

In order to find out if this `gap' is expected, we generate a stellar population
using a Monte Carlo process and two-dimensional two-sample KS tests.
For comparison with the simulated data we only consider the 59 observed 
companion and background stars with $\rho \leq 4''$. 
Objects with $\rho > 4''$ are not relevant for this test and might 
introduce biases due to observational selection effects.
We randomly draw primary and secondary masses
in the mass range $[0.02,20]~{\rm M}_\odot$.
We select only those binaries for which the primary
mass is in the range $[1.4,7.7]~{\rm M}_\odot$. 
Sets of 59 binary systems are created for 
IMF$_{-0.3}$ and IMF$_{2.5}$ (Equation~\ref{equation: imf1}).
The masses are converted into $K_S$ magnitudes 
using a distance of 145~pc and the 5~Myr
isochrone described in \S~\ref{sec: hrdiagram}. The angular separations for
the secondaries are drawn from a uniform distribution in $[0'',4'']$. We
create $10^3$ realizations which are compared to the observations.
For IMF$_{-0.3}$ we find a mean KS significance 
level of $3.5 \times 10^{-3}$, while for
IMF$_{2.5}$ we find $3.2 \times 10^{-3}$. We repeated the procedure
described above with another angular separation distribution:
$\partial N / \partial \rho \propto \rho$ for $\rho \in [0'',4'']$
(\"{O}pik's law). We find mean KS sigificance levels of
$8.2 \times 10^{-5}$ and $6.5 \times 10^{-5}$ for IMF$_{-0.3}$ and IMF$_{2.5}$, 
respectively. 
It is therefore unlikely that the observed
distribution (including the `gap') is drawn from the IMFs and angular
separation distributions as described above.

Several faint close components with $\rho \leq 3.75''$ below the `gap'
($K_S \ga 14~{\rm mag}$) are detected. These objects are found next to
the target stars 
HIP61265, HIP67260, HIP73937, HIP78968, HIP79098, HIP79410, and HIP81949.
The Strehl ratios in $K_S$ of the corresponding observations
($\sim 20\%$) are typical. The detection of these close, faint components therefore cannot be
expained by a better performance of the AO system during the observation of
these objects. 
The target stars corresponding to these objects are
not particularly faint; the luminosity contrast between the target star
and the faint close components is typical.
These objects are background stars according to our selection
criterion. However, there is a possibility that these objects are close brown dwarfs
with masses $\la 0.05~{\rm M}_\odot$, 
where the conversion from $K_S$ to mass is
strongly dependent on the age.
This result implies that A and late-B stars do not have close companions with
a mass less than about 0.1~M$_\odot$, unless the assumed background stars {\it
are} physical companions (and thus brown dwarfs). If so, a gap would be
present in the companion mass distribution.

\section{Comparison with literature data} \label{sec: literaturedata}

\subsection{New companions} \label{sec: newcompanions}

\begin{table}[bt] 
  \begin{tabular}{ll}
    \hline 
    Reference & Detection method \\
    \hline
    \cite{alencar2003} & Spectroscopic \\
    \cite{balega1994} & Visual \\
    \cite{barbier1994} & Spectroscopic\\
    \cite{batten1997} & Spectroscopic \\ 
    \cite{couteau1995} & Combination\\
    The Double Star Library & Combination\\
    \cite{duflot1995} &Spectroscopic\\
    \cite{hartkopf2001} &Visual\\ 
    \cite{jordi1997} &Eclipsing\\
    The {\it Hipparcos} and {\it Tycho} Catalogues &Astrometric\\ 
    \cite{kraicheva1989} &Spectroscopic\\ 
    \cite{malkov1993} &Combination \\
    \cite{mason1995} &Visual \\
    \cite{mcalister1993} &Visual\\
    Miscellaneous, e.g. SIMBAD &Combination\\
    \cite{miura1992} &Visual\\
    \cite{pedoussaut1996} &Spectroscopic\\ 
    \cite{shatsky2002} &Visual\\
    \cite{sowell1993} &Visual\\
    \cite{svechnikov1984} &Combination\\
    \cite{tokovinin1997} &Combination\\
    \cite{worley1997} &Combination\\
    \hline 
  \end{tabular}
  \caption{References to literature data with spectroscopic, 
    astrometric, eclipsing, and visual binaries in Sco~OB2. 
    Using these data we
    find that 33 out of the 74 candidate companion stars in our dataset were
    already documented in literature, while 41 were previously unknown.
    \label{table: litlist}}
\end{table}

We compiled a list of known companion stars to all {\it Hipparcos} members of
Sco~OB2 using literature data on binarity and multiplicity. This includes
spectroscopic, astrometric, eclipsing, and visual binaries. The references
used are listed in Table~\ref{table: litlist}.

Since the angular separation and position angle of companion stars change as a
function of time, one has to take care that a `newly' detected companion is
not a displaced known companion star. It is therefore interesting to estimate
how fast the angular separation and position angle change for the observed
companion star.
We obtained the primary mass $M_p$ and companion star mass $M_s$ 
in (\S~\ref{sec: masses}) thus we can estimate the orbital
period of the companion stars using Kepler's third law. If the orbit is
circular and the system is seen face-on, the
orbital period is given by $P = \sqrt{ 4 \pi^2 (D \rho)^3 / G (M_p+M_s) }$,
where $D$ is the distance to the system, $\rho$ the angular separation
between primary and companion star, and $G$ is the gravitational
constant. 
In general, orbits are eccentric and inclined.
However, the effect of nonzero eccentricity 
and inclination on the period is most likely
only of the order of a few per cent \citep{leinert1993}.

In our survey we are sensitive to orbits with a period
between approximately 50~yr and 50,000~yr.
For most of the systems in our sample we find an orbital period of the
order of a few
thousand years. The literature data that we used are several decades old at
most. The angular separation and position angle are not expected to have
changed significantly over this time. If a star with significantly different
values for $\rho$ and/or $\varphi$ is discovered with respect to the known
companion star, this star is considered to be new.
Several close binary systems are observed.
The shortest orbital periods that we find 
(assuming circular and face-on orbits) 
are those for
the companions of HIP62026 (57~yr), HIP62179 (68~yr), HIP76001 (84~yr),
HIP64515 (87~yr), HIP80461 (106~yr), HIP62002 (142~yr), and HIP67260
(213~yr). Differences in angular separation $\rho$ and position angle
$\varphi$ are therefore expected
with respect to previous measurements. 
For example, the observed star HIP76001 has two companions
with angular separations $\rho_1$ and $\rho_2$, and 
position angles $\varphi_1$ and $\varphi_2$, respectively.
In our AO survey we measure $(\rho_1,\varphi_1) = (0.25'',3.2^\circ)$ and
$(\rho_2,\varphi_2) = (1.48'',124.8^\circ)$, 
while \cite{tokovinin1997} quotes 
$(\rho_1,\varphi_1) = (0.094'',6^\circ)$ in 1993 and
$(\rho_2,\varphi_2) = (1.54'',130^\circ)$ in 1991.
Taking the changes in $\rho$ and $\varphi$ into account, we determine
whether an observed companion star was documented before or 
whether it is new.

\begin{table}[bt]
  \begin{tabular}{l ccccc}
    \hline \hline Type & B1-B3 & B4-B9 & A & F & GKM \\ 
    \hline \multicolumn{6}{l}{Upper Scorpius} \\ 
    \hline Single stars & 3 & 13 & 20 & 17 & 10 \\ 
    Visual & 15 & 9 & 12 & 3 & 2 \\ 
    Astrometric & 0 & 3& 1& 2& 3 \\
    Spectroscopic & 11& 12& 0 & 0& 0 \\ 
    \hline \multicolumn{6}{l}{Upper Centaurus Lupus} \\ 
    \hline 
    Single stars & 3 & 22 & 41 & 43 & 22 \\ 
    Visual  & 15 & 23 & 29 & 7 & 8 \\ 
    Astrometric & 2 & 1 & 3 & 5 & 2 \\ 
    Spectroscopic & 11& 6 & 1& 1& 0 \\ 
    \hline \multicolumn{6}{l}{Lower Centaurus Crux} \\
    \hline 
    Single stars & 3 & 16 & 31 & 47 & 14 \\ 
    Visual & 7 &13 & 24 & 10 & 4 \\ 
    Astrometric & 1 & 3 & 3 & 4 & 4 \\ 
    Spectroscopic & 3 & 3& 1 & 0 & 0 \\ 
    \hline \multicolumn{6}{l}{Scorpius OB2}\\ 
    \hline 
    Single stars & 9 & 51 & 92 & 107 & 46 \\ 
    Visual & 37 & 45 & 65 & 20 &14 \\ 
    Astrometric & 3 & 7 & 7 & 11 & 9 \\ 
    Spectroscopic & 25 & 21 & 2 & 1 & 0 \\ 
    \hline \hline
  \end{tabular}
  \caption{Binarity data for Sco~OB2, showing the number of companion stars
  found with the different techniques. Literature data and the new companion
  stars described in this paper are included. 
  The terms 'visual', 'astrometric', and 'spectroscopic' pertain only to 
  available observations and not to in intrinsic properties of the systems.
  Note that 
  each companion of a multiple system is included individually.
  \label{table: summary}}
\end{table}

Table~\ref{table: summary} summarizes the current status on binarity for
Sco~OB2. The number of presently known companion stars is listed as a function
of primary spectral type and detection method. 

We define all companions that
have only been detected with radial velocity studies as spectroscopic. All
companions of which the presence of the companion has only been derived from
astrometric studies of the primary are classified as astrometric. All other
companions, i.e. those that have been detected with (AO) imaging,
Speckle techniques, interferometry, etc., are classified as visual. Note that
our classification refers only to available observations and {\it not} to
intrinsic properties of the binary or multiple systems.

All candidate companions that are found with our AO survey are by
definition visual. 
The overlap between the observed companions in our sample and the 
spectroscopic and astrometric
binaries in literature is zero, although HIP78809 is flagged as 
``suspected non-single'' in the {\it Hipparcos} catalogue.
From our comparison with literature we find
41 new companions (14 in US; 13 in UCL; 14 in UCL), 
while 33 of the candidate companion stars were already
documented.

Several of the stars currently known as 'single' could be unresolved
systems. For a subset of the member stars, multiple companions have been
detected using different techniques. Most systems with early spectral type
primaries are found with spectroscopic methods. Amongst the A and F type
stars, most {\it Hipparcos} member stars are single, although a fraction of
these could be unresolved systems (see \S~\ref{sec: binarystatistics}).

Of the Sco~OB2 {\it Hipparcos} member stars 37 are now known to be
triple. Sixteen of these consist of a primary with two visual
companions. Fifteen consist of a primary, a visual companion, and a
spectroscopic companion. These are only found in UCL and LCC. Six consist of a
primary, a spectroscopic companion and an astrometric companion. Five
quadruple systems are known: two systems consisting of a primary and three
visual companions (HIP69113 and HIP81972) and two consisting of a primary, two
visual companions and a spectroscopic companion (HIP80112 and
HIP78384). HIP77820 is the only quintuple system, consisting of a primary,
three visual companions and one spectroscopic companion. 
The largest known system in Sco~OB2 is
HIP78374, which contains a primary, four visual companions and two
spectroscopic companions.

\subsection{Binary statistics} \label{sec: binarystatistics}

The binarity properties of a stellar population are usually quantified in
`binary fractions' \citep[e.g.,][]{reipurth1993}. 
Two common definitions in use are the multiple system
fraction $F_{\rm M}$ and non-single star fraction $F_{\rm NS}$ (since $1-F_{\rm NS}$ is the fraction
of stars that is single). Another frequently used quantity is the companion
star fraction $F_{\rm C}$\footnote{Note that in \cite{kouwenhoven2004a,kouwenhoven2004b} 
we used an incorrect definition of $F_{\rm C}$.}, which measures the average number 
of companion stars per
primary star. These quantities are defined as
\begin{eqnarray}
  F_{\rm M}       &=& (B+T+\dots) \ /\ (S+B+T+\dots);\\
  F_{\rm NS}      &=& (2B+3T+\dots) \ /\ (S+2B+3T+\dots);\\
  F_{\rm C}       &=& (B+2T+\dots) \ /\ (S+B+T+\dots),
\end{eqnarray}
where $S$, $B$, and $T$ denote the number of single systems, binary systems
and triple systems in the assocation. 
Table~\ref{table: statistics} shows several properties of the three subgroups,
including the binary statistics that are updated with our new findings.

\begin{table}[bt]
  \centering
  \setlength{\tabcolsep}{0.7\tabcolsep}
  \begin{tabular}{l cc cccc cc c}
    \hline
          & $D$  & Age  & $S$ & $B$ & $T$ & $>3$ & $F_{\rm M}$ & $F_{\rm NS}$ & $F_{\rm C}$ \\
          & (pc) &(Myr) & \\
    \hline
    US    & 145  & 5--6     & 63    & 46    & 7     & 3     & 0.47  & 0.67  & 0.61\\
    UCL   & 140  & 15--22   & 131   & 68    & 17    & 4     & 0.40  & 0.61  & 0.52\\
    LCC   & 118  & 17--23   & 112   & 54    & 13    & 0     & 0.37  & 0.57  & 0.45\\
    \hline
    all   &      &          & 303   & 171   & 37    & 7     & 0.41  & 0.61  & 0.52\\
    \hline
  \end{tabular}
  \caption{Multiplicity among {\it Hipparcos} members of Sco~OB2. The columns
  show the subgroup names (Upper Scorpius; Upper Centaurus Lupus; Lower
  Centaurus Crux), their distances \citep[see][]{dezeeuw1999}, the ages
  (\cite{degeus1989,preibisch2002} for US; \cite{mamajek2002} for UCL and LCC), the number
  of known single stars, binary stars, triple systems and $N>3$ systems, and
  the binary statistics (see \S~\ref{sec: binarystatistics}) . \label{table:
  statistics} }
\end{table}

\begin{figure}[bt]
  \centering
  \includegraphics[width=0.5\textwidth,height=!]{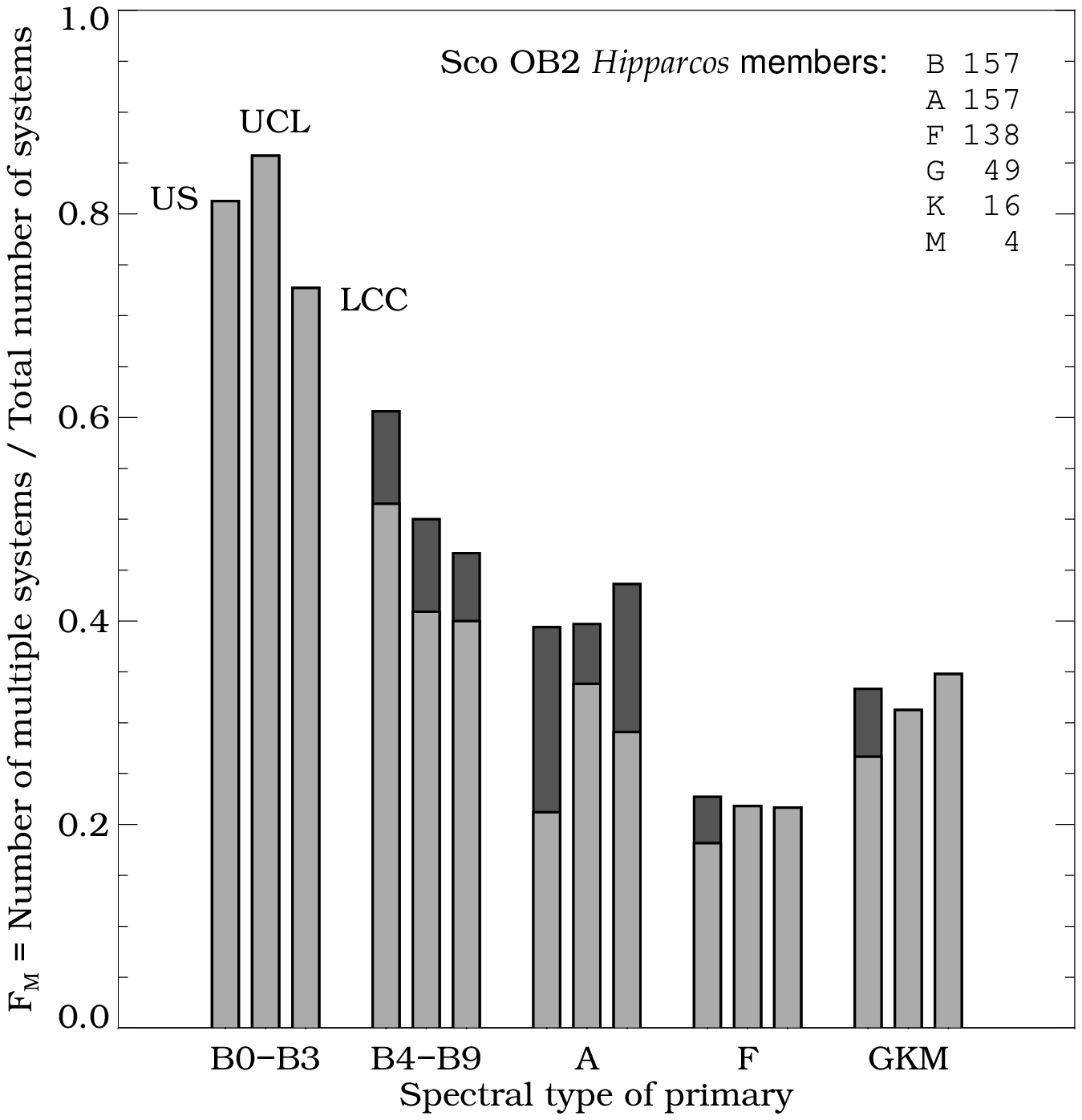}
  \caption{The fraction of stellar systems which is multiple versus the
  spectral type of the primary, for the three subgroups of Sco~OB2. Only
  confirmed Hipparcos member primaries are considered here. 
  The binarity dataset consists of literature data and
  our near-infrared AO survey.
  The light and dark
  grey parts of the bars correspond to literature data and the new data
  presented in this article, respectively. The spectral types of the companion
  stars (not included in this plot) are always later than those of the primary
  stars. Apparently, the multiplicity is a function of spectral type, but this
  conclusion may well be premature when observational biases are not properly
  taken into account. That this is at least partly true, is supported by our
  detection of 41 new close companion stars at later spectral
  types. \label{figure: mulspt}}
\end{figure}

Figure~\ref{figure: mulspt} shows the multiple system fraction $F_{\rm M}$ 
as a
function of spectral type of the primary for the three subgroups. For example,
86\% of the B0--B3 type {\it Hipparcos} member stars of UCL have one or more
known companion stars. Information about the number of companion stars and the
spectral type of the companion stars around the primary is not shown in the
figure. The new companion stars found in our AO survey are indicated with the
darker shaded parts of the bars. 
A trend between multiplicity and the spectral type
of the primary seems to be present, but this conclusion may well be premature
when observational biases are not properly taken into account. Our detection
of 41 new close companion stars shows that this is at least partially true: the
number of A-type member stars which is in a binary/multiple system in US has
doubled as a result of our survey. This strongly supports the statement made
in \cite{brown2001} that the low multiplicity for A- and F-type stars can at
least partially be explained by observational biases.

\section{Conclusions and discussion} \label{sec:conclusions}


We carried out a near-infrared adaptive optics search for companions around
199 (mainly) A and late-B stars in the nearby OB association Sco~OB2. 
Our sample is a selection of the {\it Hipparcos} membership list of Sco~OB2
in \cite{dezeeuw1999}. We find
a total of 151 stellar components around the target stars. We use a simple
brightness criterion to separate candidate companion stars ($K_S < 12~{\rm mag}$) and
probable background stars ($K_S > 12~{\rm mag}$). The validity of this criterion is
verified in several ways (\S~\ref{sec: backgroundstars}). 
Of the detected components, 77 are likely
background stars, and 74 are candidate companion stars.

The 74 candidate companions occupy the full range in $K_S$ magnitude, down to
$K_S=12~{\rm mag}$. At the distance of Sco~OB2, an M5 main sequence star has a $K_S$ magnitude
of approximately $12~{\rm mag}$. The angular separation
between primary and companion ranges from $0.22''$ to $12.4''$.
These values correspond to orbital periods of several decades to
several thousands of years.

The $J$, $H$, and $K_S$ magnitudes for all observed objects are corrected for
distance and interstellar extinction to find the absolute magnitudes $M_J$,
$M_H$, and $M_{K_S}$. 
The subset of the components with multi-color observations
are plotted in a color-magnitude diagram. 
All (except one) candidate companion stars
are positioned close to the 5~Myr (for US) and 20~Myr (for UCL and LCC)
isochrones. The background stars are positioned far away from the isochrones.

For all observed primaries and companion stars the mass is derived from the
absolute magnitude $M_{K_S}$. The mass of the A and late-B primaries ranges
between $1.4~{\rm M}_\odot$ and $7.7~{\rm M}_\odot$ and 
the mass of their secondaries between
$0.1~{\rm M}_\odot$ and $3.0~{\rm M}_\odot$. Most observed 
companion stars are less massive than
their primaries.
Most of the systems with previously undocumented companion 
stars have a mass ratio smaller than 0.25.
The mass ratio distribution for the observed objects peaks around $q=0.1$ and
decreases for higher mass ratios. The minimum and maximum value for the mass
ratio of the companion stars that we observed are $q=0.022 \pm 0.008$ 
and $1.0 \pm 0.34$. 
For systems with mass ratio $q \approx 1$ we cannot 
say with absolute certainty which is the most 
massive and therefore the primary star.
The mass ratio distribution for binaries with late-B and A type primaries follows 
the distribution $f(q)=q^{-\Gamma}$, with
$\Gamma=0.33$. This is similar to the mass ratio distribution observed 
by \cite{shatsky2002} for systems with
B type primaries. Relatively few systems with
low mass ratio are found, excluding random pairing between primaries and
companion stars.
The uncertainty in the age of the UCL and LCC subgroups does not
effect the mass ratio distribution significantly.

A cross-check with visual, astrometric and spectroscopic binaries in literature
shows that 41 of the candidate companion stars are new: 14 in US; 13 in UCL 
and 14 in LCC. The other 33 candidate companions are already documented in
literature.
We analyze the presently known data on binarity and multiplicity in Sco~OB2,
including the 41 new companions that are found in our survey. We conclude that
at least 41\% of all {\it Hipparcos} member stars of Sco~OB2 are 
either double or multiple, and we
find a companion star fraction $F_{\rm C} = 0.52$. These values are lower limits
since the dataset is affected by observational and selection biases.


Our AO observations are close to completeness in the angular
separation range $1'' \lesssim \rho \lesssim 9''$. 
Next to the 199 target stars we find 50 companion stars
in this angular separation range. This corresponds to an $F_{\rm C}$ of 0.25
in this range of angular separation 
for A and late-B type stars in Sco~OB2.
This value (not corrected for incompleteness) is slightly higher than the values for 
B type stars in Sco~OB2 
\citep[$F_{\rm C}=0.20$ per decade of $\rho$;][]{shatsky2002}
and pre-main sequence stars in 
Sco~OB2 \citep[$F_{\rm C}=0.21$ per decade of $\rho$;][]{koehler2000}.

At small angular separation ($\rho \leq 3.75''$) no components other
than the target stars are detected for magnitudes $12~{\rm mag}\lesssim K_S\lesssim
14~{\rm mag}$. This `gap' cannot be explained by observational biases or low-number
statistics. A fully populated mass spectrum for the companions (assuming
random pairing of binary components from the same underlying IMF) is also
incompatible with this gap. This implies a lower limit on the
companion masses of $\sim 0.1$~M$_\odot$. This is consistent with our finding
that the mass ratio distribution points to a deficit of low-mass companions
compared to the random pairing case.
On the other hand, if we assume that the
sources fainter than $K_S\approx14~{\rm mag}$ are actually physical companions, 
a gap may be present in the companion mass distribution.  
We will carry out follow-up multi-color AO observations to further
investigate this issue.

The gap described above might indicate a {\it brown dwarf desert},
such as observed for solar-type stars
in the solar neighbourhood \citep{duquennoy1991}.
The presence of this gap could support the embryo-ejection 
formation scenario for brown dwarfs
\citep[e.g.][]{reipurth2001,bate2003,kroupa2004}. 
This is further supported by the
detection of 28 candidate free-floating brown dwarf members of Sco~OB2 by
\cite{martin2004}.

The observed $F_{\rm M}$ decreases with decreasing primary
mass. Part of this trend was ascribed by \cite{brown2001} to observational and
selection biases. We lend support to this conclusion 
with our new AO observations. In particular, the
multiplicity fraction for A-type {\it Hipparcos} members of US is doubled.
Nevertheless our knowledge of the present day binary population in Sco~OB2 is
still rather fragmentary. None of the multiplicity surveys of Sco~OB2 so far
has been complete due to practical and time constraints, but also because of a
lack of full knowledge of the membership of the association. In addition, each
observational technique used in these surveys has its own biases. For example,
visual binaries are only detected if the angular separation between the components is
large enough with respect to the luminosity contrast, which means that with
this technique one cannot find the very short-period (spectroscopic)
binaries. Moreover the observational biases depend on the way in which a
particular survey was carried out (including how background stars were weeded
out).

We therefore intend to follow up this observational study by a careful
investigation of the effect of selection biases on the interpretation of the
results. This will be done through a detailed modeling of evolving OB
associations using state-of-the-art N-body techniques coupled with a stellar
and binary evolution code. The synthetic OB association will subsequently be
`observed' by simulating in detail the various binary surveys that have been
carried out. This modeling includes simulated photometry, adaptive optics
imaging and {\it Hipparcos} data, as well as synthetic radial velocity
surveys. Examples of this type of approach can be found in \citet{spz2001}
(for photometric data) and \citet{quist2000} (for {\it Hipparcos} data), where
the latter study very clearly demonstrates that an understanding of the
observational selection biases much enhances the interpretation of binarity
data. Finally, the time dependence of the modeling will also allow us to
investigate to what extent stellar/binary evolution and stellar dynamical effects
have altered the binary population over the lifetime of the association.

This combination of the data on the binary population in Sco OB2 and a
comprehensive modeling of both the association and the observations should
result in the most detailed description of the characteristics of the
primordial binary population to date.


\begin{acknowledgements}
We thank Francesco Palla for providing the near-infrared pre-main sequence isochrones and evolutionary tracks. We thank the anonymous referee and Andrei Tokovinin for their constructive criticism, which helped to improve the paper. This research was supported by NWO under project number 614.041.006 and the Leids Kerkhoven Bosscha Fonds.
\end{acknowledgements}

\appendix

\Online
\onecolumn
\begin{longtable}{|llllllllll|}
\caption{}\\
\hline
Star & $J$ mag & $H$ mag & $K_S$ mag & type & $\rho$ ($''$)  & $PA$  ($^\circ$)  & stat & group & remarks\\ 
\hline
\hline 
HIP50520 &  &  & 6.23 & A1V &  &  & p & LCC &  \\
 &  &  & 6.39 &  & 2.51 & 313.3 & c &  &  \\
\hline
HIP52357 &  &  & 7.64 & A3IV &  &  & p & LCC &  \\
 &  &  & 7.65 &  & 0.53 & 73.0 & c &  &  \\
 &  &  & 11.45 &  & 10.04 & 72.7 & nc &  &  \\
\hline
HIP53524 &  &  & 6.76 & A8III &  &  & p & LCC &  \\
 &  &  & 12.67 &  & 4.87 & 316.9 & b &  &  \\
\hline
HIP53701 & 6.30 & 6.37 & 6.48 & B8IV &  &  & p & LCC &  \\
 & 9.05 & 8.76 & 8.86 &  & 3.88 & 75.8 & c &  &  \\
 & 13.06 & 12.93 & 13.04 &  & 6.57 & 120.1 & b &  &  \\
\hline
HIP54231 &  &  & 6.75 & A0V &  &  & p & LCC &  \\
\hline
HIP55188 &  &  & 7.43 & A2V &  &  & p & LCC &  \\
\hline
HIP55899 &  &  & 7.07 & A0V &  &  & p & LCC &  \\
\hline
HIP56354 &  &  & 5.78 & A9V &  &  & p & LCC &  \\
\hline
HIP56379 &  &  & 5.27 & B9Vne &  &  & p & LCC &  \\
\hline
HIP56963 &  &  & 7.46 & A3V &  &  & p & LCC &  \\
\hline
HIP56993 &  &  & 7.38 & A0V &  &  & p & LCC &  \\
 &  &  & 11.88 &  & 1.68 & 23.1 & nc &  &  \\
\hline
HIP57809 &  &  & 6.61 & A0V &  &  & p & LCC &  \\
\hline
HIP58416 &  &  & 7.03 & A7V &  &  & p & LCC &  \\
 &  &  & 8.66 &  & 0.58 & 166.1 & nc &  &  \\
\hline
HIP58452 &  &  & 6.51 & B8/B9V &  &  & p & LCC &  \\
\hline
HIP58465 &  &  & 6.32 & A2V &  &  & p & LCC &  \\
\hline
HIP58720 &  &  & 6.05 & B9V &  &  & p & LCC &  \\
\hline
HIP58859 &  &  & 6.52 & B9V &  &  & p & LCC &  \\
\hline
HIP59282 &  &  & 7.00 & A3V &  &  & p & LCC &  \\
\hline
HIP59397 &  &  & 7.01 & A2V &  &  & p & LCC &  \\
\hline
HIP59413 &  &  & 7.46 & A6V &  &  & p & LCC &  \\
 &  &  & 8.18 &  & 3.18 & 99.8 & c &  &  \\
 &  &  & 15.15 &  & 7.22 & 250.3 & b &  &  \\
\hline
HIP59502 &  &  & 6.80 & A2V &  &  & p & LCC &  \\
 &  &  & 11.28 &  & 2.94 & 26.4 & nc &  &  \\
 &  &  & 12.79 &  & 9.05 & 308.6 & b &  &  \\
\hline
HIP59898 &  &  & 5.99 & A0V &  &  & p & LCC &  \\
\hline
HIP60084 &  &  & 7.65 & A1V &  &  & p & LCC &  \\
 &  &  & 10.10 &  & 0.46 & 329.6 & nc &  &  \\
\hline
HIP60183 &  &  & 6.33 & B9V &  &  & p & LCC &  \\
\hline
HIP60561 &  &  & 6.59 & A0V &  &  & p & LCC &  \\
\hline
HIP60851 &  &  & 5.98 & A0Vn &  &  & p & LCC &  \\
 &  &  & 11.04 &  & 2.01 & 44.1 & nc &  &  \\
 &  &  & 11.66 &  & 6.92 & 181.1 & nc &  &  \\
\hline
HIP61257 &  &  & 6.60 & B9V &  &  & p & LCC &  \\
 &  &  & 12.43 &  & 5.54 & 324.3 & b &  &  \\
\hline
HIP61265 &  &  & 7.44 & A2V &  &  & p & LCC &  \\
 &  &  & 11.29 &  & 2.51 & 67.4 & nc &  &  \\
 &  &  & 14.70 &  & 3.43 & 168.6 & b &  &  \\
\hline
HIP61639 &  &  & 6.94 & A1/A2V &  &  & p & LCC &  \\
 &  &  & 7.06 &  & 1.87 & 182.4 & c &  &  \\
 &  &  & 14.36 &  & 4.21 & 220.6 & b &  &  \\
\hline
HIP61782 &  &  & 7.56 & A0V &  &  & p & LCC &  \\
\hline
HIP61796 &  &  & 6.37 & B8V &  &  & p & LCC &  \\
 &  &  & 11.79 &  & 9.89 & 109.0 & nc &  &  \\
 &  &  & 11.86 &  & 12.38 & 136.8 & nc &  &  \\
 &  &  & 13.35 &  & 4.17 & 305.8 & b &  &  \\
\hline
HIP62002 &  &  & 7.09 & A1V &  &  & p & LCC &  \\
 &  &  & 7.65 &  & 0.38 & 69.2 & c &  &  \\
 &  &  & 13.95 &  & 5.33 & 214.6 & b &  &  \\
\hline
HIP62026 &  &  & 6.49 & B9V &  &  & p & LCC &  \\
 &  &  & 7.46 &  & 0.22 & 12.5 & nc &  &  \\
\hline
HIP62058 &  &  & 6.17 & B9Vn &  &  & p & LCC &  \\
\hline
HIP62179 &  &  & 7.20 & A0IV/V &  &  & p & LCC &  \\
 &  &  & 7.57 &  & 0.23 & 282.7 & c &  &  \\
 &  &  & 14.03 &  & 12.21 & 131.3 & b &  &  \\
 &  &  & 14.60 &  & 4.47 & 255.8 & b &  &  \\
\hline
HIP63204 &  &  & 6.47 & A0p &  &  & p & LCC &  \\
 &  &  & 7.36 &  & 1.80 & 46.9 & nc &  &  \\
\hline
HIP63236 &  &  & 6.66 & A2IV/V &  &  & p & LCC &  \\
 &  &  & 12.91 &  & 7.56 & 118.4 & b &  &  \\
 &  &  & 12.85 &  & 11.97 & 316.5 & b &  &  \\
\hline
HIP63839 &  &  & 6.66 & A0V &  &  & p & LCC &  \\
 &  &  & 13.78 &  & 5.99 & 358.1 & b &  &  \\
 &  &  & 14.21 &  & 4.30 & 13.7 & b &  &  \\
 &  &  & 13.16 &  & 6.31 & 300.9 & b &  &  \\
\hline
HIP64320 &  &  & 6.22 & Ap &  &  & p & LCC &  \\
\hline
HIP64515 &  &  & 6.78 & B9V &  &  & p & LCC &  \\
 &  &  & 6.94 &  & 0.31 & 165.7 & c &  &  \\
\hline
HIP64892 &  &  & 6.82 & B9V &  &  & p & LCC &  \\
\hline
HIP64925 &  &  & 6.88 & A0V &  &  & p & LCC &  \\
\hline
HIP64933 &  &  & 6.29 & A0V &  &  & p & LCC &  \\
\hline
HIP65021 &  &  & 7.26 & B9V &  &  & p & LCC &  \\
\hline
HIP65089 &  &  & 7.37 & A7/A8V &  &  & p & LCC &  \\
\hline
HIP65178 &  &  & 6.71 & B9V &  &  & p & LCC &  \\
\hline
HIP65219 &  &  & 6.52 & A3/A4III/IV &  &  & p & LCC &  \\
\hline
HIP65394 &  &  & 7.25 & A1Vn... &  &  & p & LCC &  \\
\hline
HIP65426 &  &  & 6.78 & A2V &  &  & p & LCC &  \\
\hline
HIP65822 &  &  & 6.68 & A1V &  &  & p & LCC &  \\
 &  &  & 11.08 &  & 1.82 & 303.9 & nc &  &  \\
\hline
HIP65965 &  &  & 7.51 & B9V &  &  & p & LCC &  \\
 &  &  & 15.21 &  & 10.28 & 41.1 & b &  &  \\
\hline
HIP66068 &  &  & 7.04 & A1/A2V &  &  & p & LCC &  \\
\hline
HIP66447 &  &  & 7.16 & A3IV/V &  &  & p & UCL &  \\
\hline
HIP66454 &  &  & 6.33 & B8V &  &  & p & LCC &  \\
\hline
HIP66566 &  &  & 7.36 & A1V &  &  & p & LCC &  \\
\hline
HIP66651 &  &  & 7.35 & B9.5V &  &  & p & LCC &  \\
 &  &  & 15.36 &  & 7.58 & 173.3 & b &  &  \\
\hline
HIP66722 &  &  & 6.32 & A0V &  &  & p & UCL &  \\
\hline
HIP66908 &  &  & 6.86 & A4V &  &  & p & UCL &  \\
\hline
HIP67036 &  &  & 6.69 & A0p &  &  & p & LCC &  \\
\hline
HIP67260 &  &  & 7.03 & A0V &  &  & p & LCC &  \\
 &  &  & 8.37 &  & 0.44 & 228.9 & c &  &  \\
 &  &  & 14.70 &  & 2.22 & 77.1 & b &  &  \\
\hline
HIP67919 &  &  & 6.58 & A9V &  &  & p & LCC &  \\
 &  &  & 9.05 &  & 0.70 & 299.1 & nc &  &  \\
\hline
HIP68080 &  &  & 6.28 & A1V &  &  & p & UCL &  \\
 &  &  & 7.19 &  & 1.92 & 10.2 & c &  &  \\
\hline
HIP68532 &  &  & 7.03 & A3IV/V &  &  & p & UCL &  \\
 &  &  & 9.50 &  & 3.03 & 288.6 & nc &  &  \\
 &  &  & 10.53 &  & 3.15 & 291.7 & nc &  &  \\
\hline
HIP68781 &  &  & 7.38 & A2V &  &  & p & UCL &  \\
\hline
HIP68867 &  &  & 7.17 & A0V &  &  & p & UCL &  \\
 &  &  & 11.61 &  & 2.16 & 284.8 & nc &  &  \\
\hline
HIP68958 &  &  & 6.72 & Ap... &  &  & p & UCL &  \\
\hline
HIP69113 & 6.25 & 6.32 & 6.43 & B9V &  &  & p & UCL &  \\
 & 11.14 & 10.51 & 10.25 &  & 5.33 & 64.8 & nc &  &  \\
 & 11.19 & 10.46 & 10.23 &  & 5.52 & 66.9 & nc &  &  \\
\hline
HIP69749 &  &  & 6.62 & B9IV &  &  & p & UCL &  \\
 &  &  & 11.60 &  & 1.50 & 0.8 & nc &  &  \\
 &  &  & 12.69 &  & 8.10 & 50.0 & b &  &  \\
 &  &  & 13.40 &  & 5.57 & 352.4 & b &  &  \\
 &  &  & 14.16 &  & 8.79 & 62.5 & b &  &  \\
 &  &  & 14.48 &  & 9.13 & 66.8 & b &  &  \\
 &  &  & 14.51 &  & 4.43 & 278.6 & b &  &  \\
\hline
HIP69845 &  &  & 7.78 & B9V &  &  & p & UCL &  \\
\hline
HIP70441 &  &  & 7.31 & A1V &  &  & p & UCL &  \\
\hline
HIP70455 &  &  & 7.07 & B8V &  &  & p & UCL &  \\
\hline
HIP70626 &  &  & 6.56 & B9V &  &  & p & UCL &  \\
\hline
HIP70690 &  &  & 7.71 & B9V &  &  & p & UCL &  \\
\hline
HIP70697 &  &  & 7.17 & A0V &  &  & p & UCL &  \\
\hline
HIP70809 &  &  & 6.54 & Ap... &  &  & p & UCL &  \\
 &  &  & 14.64 &  & 4.97 & 214.4 & b &  &  \\
 &  &  & 14.60 &  & 8.72 & 297.3 & b &  &  \\
\hline
HIP70904 &  &  & 6.39 & A6V &  &  & p & UCL &  \\
 &  &  & 12.08 &  & 6.08 & 120.3 & b &  &  \\
 &  &  & 14.17 &  & 10.03 & 309.2 & b &  &  \\
\hline
HIP70918 &  &  & 6.35 & A0/A1V &  &  & p & UCL &  \\
\hline
HIP70998 &  &  & 7.06 & A1V &  &  & p & UCL &  \\
 &  &  & 10.83 &  & 1.17 & 354.6 & nc &  &  \\
\hline
HIP71140 &  &  & 7.13 & A7/A8IV &  &  & p & UCL &  \\
\hline
HIP71271 &  &  & 7.57 & A0V &  &  & p & UCL &  \\
\hline
HIP71321 &  &  & 7.17 & A9V &  &  & p & UCL &  \\
\hline
HIP71724 &  &  & 6.79 & B8/B9V &  &  & p & UCL &  \\
 &  &  & 9.70 &  & 8.66 & 23.0 & c &  &  \\
\hline
HIP71727 &  &  & 6.89 & A0p &  &  & p & UCL &  \\
 &  &  & 7.80 &  & 9.14 & 245.0 & c &  &  \\
\hline
HIP72140 &  &  & 7.09 & A1IV/V &  &  & p & UCL &  \\
 &  &  & 12.21 &  & 4.51 & 229.1 & b &  &  \\
\hline
HIP72192 & 6.71 & 6.71 & 6.71 & A0V &  &  & p & UCL &  \\
\hline
HIP72627 & 6.54 & 6.54 & 6.53 & A2V &  &  & p & UCL &  \\
\hline
HIP72940 &  &  & 6.85 & A1V &  &  & p & UCL &  \\
 &  &  & 8.57 &  & 3.16 & 221.6 & c &  &  \\
\hline
HIP72984 &  &  & 7.05 & A0/A1V &  &  & p & UCL &  \\
 &  &  & 14.37 &  & 5.83 & 118.1 & b &  &  \\
 &  &  & 8.50 &  & 4.71 & 260.3 & c &  &  \\
\hline
HIP73145 &  &  & 7.54 & A2IV &  &  & p & UCL &  \\
\hline
HIP73266 &  &  & 7.30 & B9V &  &  & p & UCL &  \\
\hline
HIP73341 &  &  & 6.67 & B8V &  &  & p & UCL &  \\
\hline
HIP73393 &  &  & 7.21 & A0V &  &  & p & UCL &  \\
\hline
HIP73937 &  &  & 6.05 & Ap &  &  & p & UCL &  \\
 &  &  & 14.05 &  & 3.48 & 30.5 & b &  &  \\
\hline
HIP74066 &  &  & 6.08 & B8IV &  &  & p & UCL &  \\
 &  &  & 8.43 &  & 1.22 & 109.6 & nc &  &  \\
\hline
HIP74100 &  &  & 6.12 & B7V &  &  & p & UCL &  \\
\hline
HIP74479 &  &  & 6.31 & B8V &  &  & p & UCL &  \\
 &  &  & 10.83 &  & 4.65 & 154.1 & nc &  &  \\
\hline
HIP74657 &  &  & 6.97 & B9IV &  &  & p & UCL &  \\
\hline
HIP74752 &  &  & 6.84 & B9V &  &  & p & UCL &  \\
 &  &  & 13.13 &  & 9.66 & 21.0 & b &  &  \\
\hline
HIP74797 &  &  & 7.55 & A2IV &  &  & p & UCL &  \\
\hline
HIP74985 &  &  & 7.53 & A0V &  &  & p & UCL &  \\
 &  &  & 13.13 &  & 6.37 & 145.2 & b &  &  \\
\hline
HIP75056 &  &  & 7.31 & A2V &  &  & p & UCL &  \\
 &  &  & 11.17 &  & 5.19 & 34.5 & nc &  &  \\
\hline
HIP75077 &  &  & 6.97 & A1V &  &  & p & UCL &  \\
\hline
HIP75151 &  &  & 6.65 & A+... &  &  & p & UCL &  \\
 &  &  & 8.09 &  & 5.70 & 120.9 & c &  &  \\
\hline
HIP75210 &  &  & 6.82 & B8/B9V &  &  & p & UCL &  \\
\hline
HIP75476 &  &  & 6.88 & A1/A2V &  &  & p & UCL &  \\
\hline
HIP75509 &  &  & 7.40 & A2V &  &  & p & UCL &  \\
\hline
HIP75647 &  &  & 5.86 & B5V &  &  & p & UCL &  \\
\hline
HIP75915 &  &  & 6.44 & B9V &  &  & p & UCL &  \\
 &  &  & 8.15 &  & 5.60 & 229.4 & c &  &  \\
\hline
HIP75957 &  &  & 7.24 & A0V &  &  & p & UCL &  \\
 &  &  & 13.41 &  & 5.56 & 105.7 & b &  &  \\
 &  &  & 13.21 &  & 9.21 & 227.1 & b &  &  \\
\hline
HIP76001 &  &  & 7.60 & A2/A3V &  &  & p & UCL &  \\
 &  &  & 7.80 &  & 0.25 & 3.2 & c &  &  \\
 &  &  & 8.20 &  & 1.48 & 124.8 & c &  &  \\
 &  &  & 12.85 &  & 6.58 & 127.5 & b &  &  \\
\hline
HIP76048 &  &  & 6.26 & B6/B7V &  &  & p & UCL &  \\
\hline
HIP76071 & 7.05 & 7.10 & 7.06 & B9V &  &  & p & US &  \\
 &  & 11.28 & 10.87 &  & 0.69 & 40.8 & nc &  &  \\
\hline
HIP76310 &  &  & 7.35 & A0V &  &  & p & US &  \\
\hline
HIP76503 &  &  & 6.25 & B9IV &  &  & p & US &  \\
\hline
HIP76633 &  &  & 7.51 & B9V &  &  & p & US &  \\
\hline
HIP77150 &  &  & 7.28 & A2V &  &  & p & UCL &  \\
\hline
HIP77295 &  &  & 7.64 & A2IV/V &  &  & p & UCL &  \\
 &  &  & 15.13 &  & 4.63 & 309.9 & b &  &  \\
\hline
HIP77315 &  &  & 7.24 & A0V   &       &       & p & UCL &  \\
 &  &  & 7.92 &  & 0.68 & 67.0 & c &  &  \\
HIP77317 &  &  & 7.37 & B9.5V & 37.37 & 137.3 & p & UCL &  \\
 &  &  & 14.24 &  & 32.78 & 146.8 & b &  &  \\
\hline
%
HIP77457 &  &  & 7.33 & A7IV &  &  & p & US &  \\
\hline
HIP77523 &  &  & 7.41 & B9V &  &  & p & UCL &  \\
\hline
HIP77858 &  &  & 5.40 & B5V &  &  & p & US &  \\
\hline
HIP77859 &  &  & 5.57 & B2V &  &  & p & US &  \\
\hline
HIP77900 &  &  & 6.44 & B7V &  &  & p & US &  \\
\hline
HIP77909 &  &  & 6.19 & B8III/IV &  &  & p & US &  \\
\hline
HIP77911 & 6.67 & 6.71 & 6.68 & B9V &  &  & p & US &  \\
 & 12.68 & 12.20 & 11.84 &  & 7.96 & 279.3 & nc &  &  \\
\hline
HIP77939 &  &  & 6.56 & B2/B3V &  &  & p & US &  \\
 &  &  & 8.09 &  & 0.52 & 119.1 & c &  &  \\
\hline
HIP77968 &  &  & 7.00 & B8V &  &  & p & UCL &  \\
 &  &  & 12.64 &  & 6.54 & 344.4 & b &  &  \\
 &  &  & 14.56 &  & 6.43 & 349.8 & b &  &  \\
\hline
HIP78099 &  &  & 7.35 & A0V &  &  & p & US &  \\
\hline
HIP78168 &  &  & 5.91 & B3V &  &  & p & US &  \\
\hline
HIP78196 &  &  & 7.08 & A0V &  &  & p & US &  \\
\hline
HIP78246 &  &  & 5.84 & B5V &  &  & p & US &  \\
\hline
HIP78494 &  &  & 7.11 & A2m... &  &  & p & US &  \\
\hline
HIP78530 & 6.87 & 6.92 & 6.87 & B9V &  &  & p & US &  \\
 &  & 14.56 & 14.22 &  & 4.54 & 139.7 & b &  &  \\
\hline
HIP78533 &  &  & 6.99 & Ap &  &  & p & UCL &  \\
 &  &  & 12.28 &  & 6.09 & 186.3 & b &  &  \\
\hline
HIP78541 &  &  & 6.99 & A0V &  &  & p & UCL &  \\
\hline
HIP78549 &  &  & 7.13 & B9.5V &  &  & p & US &  \\
 &  &  & 14.62 &  & 11.78 & 47.3 & b &  &  \\
\hline
HIP78663 &  &  & 7.76 & F5V &  &  & p & US &  \\
 &  &  & 13.42 &  & 8.88 & 103.0 & b &  &  \\
 &  &  & 15.41 &  & 6.11 & 184.7 & b &  &  \\
\hline
HIP78702 &  &  & 7.41 & B9V &  &  & p & US &  \\
\hline
HIP78754 &  &  & 6.95 & B8/B9V &  &  & p & UCL &  \\
\hline
HIP78756 &  &  & 7.16 & Ap &  &  & p & UCL &  \\
 &  &  & 9.52 &  & 8.63 & 216.4 & c &  &  \\
\hline
HIP78809 & 7.41 & 7.50 & 7.51 & B9V &  &  & p & US &  \\
 & 11.08 & 10.45 & 10.26 &  & 1.18 & 25.7 & nc &  &  \\
\hline
HIP78847 &  &  & 7.32 & A0V &  &  & p & US &  \\
 &  &  & 11.30 &  & 8.95 & 164.0 & nc &  &  \\
\hline
HIP78853 &  &  & 7.50 & A5V &  &  & p & UCL &  \\
 &  &  & 8.45 &  & 1.99 & 270.4 & c &  &  \\
 &  &  & 15.02 &  & 7.12 & 84.2 & b &  &  \\
\hline
HIP78877 &  &  & 6.08 & B8V &  &  & p & US &  \\
\hline
HIP78956 & 7.52 & 7.54 & 7.57 & B9.5V &  &  & p & US &  \\
 & 9.76 & 9.12 & 9.04 &  & 1.02 & 48.7 & nc &  &  \\
\hline
HIP78968 & 7.40 & 7.41 & 7.47 & B9V &  &  & p & US &  \\
 &  & 14.59 & 14.47 &  & 2.75 & 321.1 & b &  &  \\
\hline
HIP78996 &  &  & 7.46 & A9V &  &  & p & US &  \\
\hline
HIP79031 &  &  & 7.00 & B8IV/V &  &  & p & US &  \\
\hline
HIP79044 &  &  & 6.91 & B9V &  &  & p & UCL &  \\
 &  &  & 15.18 &  & 5.02 & 91.6 & b &  &  \\
\hline
HIP79098 &  &  & 5.90 & B9V &  &  & p & US &  \\
 &  &  & 14.10 &  & 2.37 & 116.8 & b &  &  \\
\hline
HIP79124 & 7.16 & 7.14 & 7.13 & A0V &  &  & p & US &  \\
 & 11.38 & 10.55 & 10.38 &  & 1.02 & 96.2 & nc &  &  \\
\hline
HIP79156 & 7.56 & 7.56 & 7.61 & A0V &  &  & p & US &  \\
 & 11.62 & 10.89 & 10.77 &  & 0.89 & 58.9 & nc &  &  \\
\hline
HIP79250 &  &  & 7.49 & A3III/IV &  &  & p & US &  \\
 &  &  & 10.71 &  & 0.62 & 180.9 & nc &  &  \\
\hline
HIP79366 &  &  & 7.47 & A3V &  &  & p & US &  \\
\hline
HIP79410 &  &  & 7.05 & B9V &  &  & p & US &  \\
 &  &  & 14.93 &  & 3.17 & 339.8 & b &  &  \\
\hline
HIP79439 &  &  & 6.97 & B9V &  &  & p & US &  \\
\hline
HIP79530 &  &  & 6.60 & B6IV &  &  & p & US &  \\
 &  &  & 8.34 &  & 1.69 & 219.7 & c &  &  \\
\hline
HIP79599 &  &  & 6.30 & B9V &  &  & p & US &  \\
\hline
HIP79622 &  &  & 6.34 & B8V &  &  & p & US &  \\
\hline
HIP79631 &  &  & 7.17 & B9.5V &  &  & p & UCL &  \\
 &  &  & 7.61 &  & 2.94 & 127.9 & nc &  &  \\
 &  &  & 14.08 &  & 8.86 & 151.8 & b &  &  \\
\hline
HIP79739 &  &  & 7.24 & B8V &  &  & p & US &  \\
 &  &  & 11.60 &  & 0.94 & 118.6 & nc &  &  \\
\hline
HIP79771 &  &  & 7.09 & B9V &  &  & p & US &  \\
 &  &  & 10.94 &  & 3.66 & 313.6 & nc &  &  \\
\hline
HIP79785 &  &  & 6.41 & B9V &  &  & p & US &  \\
\hline
HIP79860 &  &  & 7.88 & A0V &  &  & p & US &  \\
\hline
HIP79878 &  &  & 7.06 & A0V &  &  & p & US &  \\
\hline
HIP79897 &  &  & 6.99 & B9V &  &  & p & US &  \\
\hline
HIP80019 &  &  & 7.08 & A0V &  &  & p & US &  \\
\hline
HIP80024 &  &  & 6.73 & B9II/III &  &  & p & US &  \\
\hline
HIP80059 &  &  & 7.44 & A7III/IV &  &  & p & US &  \\
\hline
HIP80126 &  &  & 6.44 & B6/B7Vn &  &  & p & US &  \\
\hline
HIP80142 &  &  & 6.60 & B7V &  &  & p & UCL &  \\
 &  &  & 12.16 &  & 8.54 & 44.0 & b &  &  \\
 &  &  & 9.53 &  & 9.32 & 216.2 & c &  &  \\
\hline
HIP80238 & 7.45 & 7.45 & 7.34 & A1III/IV &  &  & p & US &  \\
 & 7.96 & 7.66 & 7.49 &  & 1.03 & 318.5 & c &  &  \\
\hline
HIP80324 &  &  & 7.33 & A0V+... &  &  & p & US &  \\
 &  &  & 7.52 &  & 6.23 & 152.5 & c &  &  \\
\hline
HIP80371 &  &  & 6.40 & B5III &  &  & p & US &  \\
 &  &  & 8.92 &  & 2.73 & 140.6 & c &  &  \\
 &  &  & 13.36 &  & 9.22 & 32.0 & b &  &  \\
\hline
HIP80425 &  &  & 7.40 & A1V &  &  & p & US &  \\
 &  &  & 8.63 &  & 0.60 & 155.8 & c &  &  \\
\hline
HIP80461 &  &  & 5.92 & B3/B4V &  &  & p & US &  \\
 &  &  & 7.09 &  & 0.27 & 285.6 & c &  &  \\
\hline
HIP80474 & 6.13 & 6.08 & 5.76 & A &  &  & p & US & JH \\
 & 13.00 & 12.30 & 11.71 &  & 4.83 & 206.2 & nc &  & JH \\
\hline
HIP80493 &  &  & 7.05 & B9V &  &  & p & US &  \\
\hline
HIP80591 &  &  & 7.82 & A5V &  &  & p & UCL &  \\
\hline
HIP80799 & 7.68 & 7.75 & 7.46 & A2V &  &  & p & US & JH \\
 & 10.94 & 10.35 & 9.88 &  & 2.94 & 205.2 & nc &  & JH \\
\hline
HIP80896 &  &  & 7.53 & F3V &  &  & p & US &  \\
 &  &  & 10.51 &  & 2.28 & 177.0 & nc &  &  \\
\hline
HIP80897 &  &  & 7.78 & A0V &  &  & p & UCL &  \\
\hline
HIP81136 &  &  & 5.21 & A7/A8+... &  &  & p & UCL &  \\
\hline
HIP81316 &  &  & 6.72 & B9V &  &  & p & UCL &  \\
\hline
HIP81472 &  &  & 5.96 & B2.5IV &  &  & p & UCL &  \\
 &  &  & 13.23 &  & 5.21 & 357.5 & b &  &  \\
 &  &  & 13.20 &  & 4.52 & 274.4 & b &  &  \\
\hline
HIP81474 &  &  & 5.74 & B9.5IV &  &  & p & US &  \\
\hline
HIP81624 &  &  & 5.80 & A1V &  &  & p & US &  \\
 &  &  & 7.95 &  & 1.13 & 224.3 & c &  &  \\
\hline
HIP81751 &  &  & 8.29 & A9V &  &  & p & UCL &  \\
 &  &  & 12.18 &  & 8.91 & 68.0 & b &  &  \\
 &  &  & 14.44 &  & 6.12 & 219.6 & b &  &  \\
\hline
HIP81914 &  &  & 6.34 & B6/B7V &  &  & p & UCL &  \\
 &  &  & 12.29 &  & 6.18 & 49.5 & b &  &  \\
 &  &  & 14.28 &  & 9.11 & 39.8 & b &  &  \\
 &  &  & 14.61 &  & 5.15 & 286.2 & b &  &  \\
\hline
HIP81949 &  &  & 7.32 & A3V &  &  & p & UCL &  \\
 &  &  & 13.23 &  & 3.90 & 89.5 & b &  &  \\
 &  &  & 14.05 &  & 3.42 & 28.7 & b &  &  \\
 &  &  & 14.70 &  & 9.60 & 76.6 & b &  &  \\
 &  &  & 14.63 &  & 6.31 & 239.0 & b &  &  \\
 &  &  & 14.62 &  & 5.69 & 292.2 & b &  &  \\
 &  &  & 15.15 &  & 5.22 & 340.5 & b &  &  \\
 &  &  & 15.33 &  & 9.70 & 345.5 & b &  &  \\
\hline
HIP81972 &  &  & 5.92 & B3V &  &  & p & UCL &  \\
 &  &  & 10.57 &  & 2.01 & 312.0 & nc &  &  \\
 &  &  & 10.75 &  & 7.07 & 258.4 & c &  &  \\
 &  &  & 11.54 &  & 5.03 & 213.7 & nc &  &  \\
\hline
HIP82154 & 6.89 & 7.22 & 7.05 & B9IV/V &  &  & p & UCL & JH \\
 & 14.86 & 14.93 & 14.47 &  & 8.39 & 359.0 & b &  & JH \\
\hline
HIP82397 &  &  & 7.28 & A3V &  &  & p & US &  \\
 &  &  & 15.36 &  & 7.88 & 227.7 & b &  &  \\
\hline
HIP82430 &  &  & 7.25 & B9V &  &  & p & UCL &  \\
 &  &  & 12.41 &  & 4.59 & 96.1 & b &  &  \\
 &  &  & 14.03 &  & 5.98 & 65.3 & b &  &  \\
 &  &  & 14.29 &  & 6.08 & 329.3 & b &  &  \\
\hline
HIP82560 & 6.83 & 6.97 & 6.58 & A0V &  &  & p & UCL & JH \\
 &  & 13.86 & 12.76 &  & 4.73 & 4.6 & b &  & H \\
 &  & 14.40 & 13.13 &  & 3.94 & 222.0 & b &  & H \\
 &  & 14.38 & 13.68 &  & 6.20 & 283.3 & b &  & H \\
\hline
HIP83457 &  &  & 6.49 & A9V &  &  & p & UCL &  \\
\hline
HIP83542 &  &  & 5.38 & G8/K0III &  &  & p & US &  \\
 &  &  & 10.01 &  & 8.96 & 196.1 & nc &  &  \\
\hline
HIP83693 &  &  & 5.69 & A2IV &  &  & p & UCL &  \\
 &  &  & 9.26 &  & 5.82 & 78.4 & c &  &  \\
 &  &  & 13.64 &  & 12.69 & 134.9 & b &  &  \\
\hline

\end{longtable}


\begin{thebibliography}{}

\bibitem[Alencar et al.(2003)]{alencar2003} Alencar, S.~H.~P., 
Melo, C.~H.~F., Dullemond, C.~P., Andersen, J., Batalha, C., Vaz, L.~P.~R., 
\& Mathieu, R.~D.\ 2003, \aap, 409, 1037 

\bibitem[Balega et al.(1994)]{balega1994} Balega, I.~I., Balega, 
Y.~Y., Belkin, I.~N., Maximov, A.~F., Orlov, V.~G., Pluzhnik, E.~A., 
Shkhagosheva, Z.~U., \& Vasyuk, V.~A.\ 1994, \aaps, 105, 503 

\bibitem[Baraffe et al.(1998)]{baraffe1998} Baraffe, I., Chabrier, G., Allard, 
F., \& Hauschildt, P.~H.\ 1998, \aap, 337, 403 

\bibitem[Barbier-Brossat, Petit, \& Figon(1994)]{barbier1994} 
Barbier-Brossat, M., Petit, M., \& Figon, P.\ 1994, \aaps, 108, 603 

\bibitem[Bate, Bonnell, \& Bromm(2003)]{bate2003} Bate, M.~R., 
Bonnell, I.~A., \& Bromm, V.\ 2003, \mnras, 339, 577 

\bibitem[Batten, Fletcher, \& MacCarthy(1997)]{batten1997} Batten, 
A.~H., Fletcher, J.~M., \& MacCarthy, D.~G.\ 1997, VizieR Online Data 
Catalog, 5064,  

\bibitem[Bertelli et al.(1994)]{bertelli1994} Bertelli, G., Bressan, 
A., Chiosi, C., Fagotto, F., \& Nasi, E.\ 1994, \aaps, 106, 275 

\bibitem[Beuzit et al.(1997)]{beuzit1997} Beuzit, J.-L.~et al.\ 
1997, Experimental Astronomy, 7, 285 

\bibitem[Blaauw(1964)]{blaauw1964} Blaauw, A.\ 1964, \araa, 2, 213 

\bibitem[Blaauw(1978)]{blaauw1978} Blaauw, A.\ 1978, Problems of 
Physics and Evolution of the Universe, 101 

\bibitem[Blaauw(1991)]{blaauw1991} 
Blaauw, A.\ 1991, in NATO ASI Ser.\ C Vol.\ 342, The Physics of Star Formation
and Early Stellar Evolution, ed C.J.\ Lada \& N.D.\ Kylafis, (Dordrecht:
Kluwer), 125

\bibitem[van der Bliek, Manfroid, \& Bouchet(1996)]{bliek1996} 
van der Bliek, N.~S., Manfroid, J., \& Bouchet, P.\ 1996, \aaps, 119, 547 




 
\bibitem[Brown et~al.(1999)]{brown1999}
Brown, A.~G.~A., Blaauw, A., Hoogerwerf, R., de Bruijne, J.~H.~J., de Zeeuw,
P.~T.\ 1999, in NATO ASI Ser.\ C Vol.\ 540, The Origin of Stars and Planetary
Systems, ed C.J.\ Lada \& N.D.\ Kylafis, (Dordrecht: Kluwer), p.\ 411

\bibitem[Brown(2001)]{brown2001} Brown, A.\ 2001, Astronomische 
Nachrichten, 322, 43 

\bibitem[de Bruijne(1999)]{debruijne1999} de Bruijne, J.~H.~J.\ 1999, 
\mnras, 310, 585 

\bibitem[Carpenter(2001)]{carpenter2001} Carpenter, J.~M.\ 2001, \aj, 
121, 2851 

\bibitem[Carter \& Meadows(1995)]{carter1995} Carter, B.~S.~\& 
Meadows, V.~S.\ 1995, \mnras, 276, 734 

\bibitem[Chabrier et al.(2000)]{chabrier2000} Chabrier, G., Baraffe, I., Allard, 
F., \& Hauschildt, P.\ 2000, \apj, 542, 464 

\bibitem[Couteau(1995)]{couteau1995} Couteau, P.\ 1995, VizieR 
Online Data Catalog, 1209 

\bibitem[Devillard(1997)]{eclipseref} Devillard, N. \ 1997, 
"The eclipse software", The Messenger No 87, 19

\bibitem[Diolaiti et al.(2000)]{diolaiti2000} Diolaiti, E., 
Bendinelli, O., Bonaccini, D., Close, L., Currie, D., \& Parmeggiani, G.\ 
2000, \aaps, 147, 335 

\bibitem[Duflot, Figon, \& Meyssonnier(1995)]{duflot1995} Duflot, 
M., Figon, P., \& Meyssonnier, N.\ 1995, \aaps, 114, 269 

\bibitem[Duquennoy \& Mayor(1991)]{duquennoy1991} Duquennoy, A.~\& 
Mayor, M.\ 1991, \aap, 248, 485 

\bibitem[ESA(1997)]{ESA97}
ESA\ 1997, The Hipparcos and Tycho Catalogues, ESA SP--1200

\bibitem[de Geus, de Zeeuw, \& Lub(1989)]{degeus1989} de Geus, 
E.~J., de Zeeuw, P.~T., \& Lub, J.\ 1989, \aap, 216, 44 

\bibitem[Girardi et al.(2002)]{girardi2002} Girardi, L., Bertelli, 
G., Bressan, A., Chiosi, C., Groenewegen, M.~A.~T., Marigo, P., Salasnich, 
B., \& Weiss, A.\ 2002, \aap, 391, 195 

\bibitem[Hartkopf, McAlister, \& Mason(2001)]{hartkopf2001} 
Hartkopf, W.~I., McAlister, H.~A., \& Mason, B.~D.\ 2001, \aj, 122, 3480 

\bibitem[Hoogerwerf(2000)]{hoogerwerf2000}
Hoogerwerf, R.\ 2000, MNRAS, 313, 43

\bibitem[Hogeveen(1990)]{hogeveen1990} Hogeveen, S.~J.\ 1990, \apss, 
173, 315 

\bibitem[Hu{\' e}lamo et al.(2001)]{huelamo2001} Hu{\' e}lamo, N., 
Brandner, W., Brown, A.~G.~A., Neuh{\" a}user, R., \& Zinnecker, H.\ 2001, 
\aap, 373, 657 

\bibitem[Hut et al.(2003)]{hut2003}
Hut, P.~et al.\ 2003, New Astronomy, 8, 337

\bibitem[Jordi et al.(1997)]{jordi1997} Jordi, 
C., Ribas, I., Torra, J., \& Gimenez, A.\ 1997, \aap, 326, 1044 

\bibitem[Kenyon \& Hartmann(1995)]{kenyon1995} Kenyon, S.~J.~\& 
Hartmann, L.\ 1995, \apjs, 101, 117 

\bibitem[Klessen, Heitsch, \& Mac Low(2000)]{klessen2000}
Klessen, R.~S., Heitsch, F., \& Mac Low, M.\ 2000, \apj, 535, 887 

\bibitem[K\"ohler et al.(2000)]{koehler2000}
K\"ohler, R., Kunkel, M., Leinert, Ch., Zinnecker, H.\ 2000, \aap, 356, 541

\bibitem[Kouwenhoven et al.(2004a)]{kouwenhoven2004a}
Kouwenhoven, M.~B.~N., Brown, A.~G.~A., Gualandris, A., Kaper, L., Portegies
Zwart, S.~F., Zinnecker, H. \ 2004, to appear in proceedings of IAU Coll. 191 'The environments and evolution of binary and multiple stars', held in M\'erida, Mexico, 3-7 February 2003, Edited by Allen,~C. \& Scarfe,~C.

\bibitem[Kouwenhoven et al.(2004b)]{kouwenhoven2004b}
Kouwenhoven, M.~B.~N., Brown, A.~G.~A., Zinnecker, H. Kaper, L., 
Portegies, Zwart, S.~F., Gualandris, A. \ 2004, 
to appear in the proceedings of the ESO Workshop on 'Science with Adaptive Optics', held in Garching, Germany, 16-19 September 2003, Edited by Brandner,~W. \& Kasper,~M.

\bibitem[Kraicheva et al.(1989)]{kraicheva1989} Kraicheva, Z.~T., Popova, E.~I., 
Tutukov, A.~V., \& Yungelson, L.~R.\ 1989, Nauchnye Informatsii, 67, 3 

\bibitem[Kroupa(1995a)]{kroupa1995a}
Kroupa, P.\ 1995a, \mnras, 277, 1492

\bibitem[Kroupa(1995b)]{kroupa1995b}
Kroupa, P.\ 1995b, \mnras, 277, 1507

\bibitem[Kroupa(1995c)]{kroupa1995c}
Kroupa, P.\ 1995c, \mnras, 277, 1522

\bibitem[Kroupa, Aarseth, \& Hurley(2001)]{kroupa2001}
Kroupa, P., Aarseth, S., \& Hurley, J.\ 2001, \mnras, 321, 699

\bibitem[Kroupa(2002)]{kroupa2002} Kroupa, P.\ 2002, Science, 295, 
82 

\bibitem[Kroupa \& Boily(2002)]{kroupaboily2002} Kroupa, P.~\& 
Boily, C.~M.\ 2002, \mnras, 366, 1188

\bibitem[Kroupa \& Bouvier(2003)]{kroupa2004} Kroupa, P.~\& 
Bouvier, J.\ 2003, \mnras, 346, 369 

\bibitem[Lada \& Lada(2003)]{LL2003}
Lada, C.~J., Lada, E.~A.\ 2003, \araa, 41, 57

\bibitem[Larson(2001)]{larson2001}
Larson, R.~B.\ 2001, in IAU Symp.\ 200, The Formation of Binary Stars, ed.\
H.\ Zinnecker and R.~D.\ Mathieu (San Fransisco: ASP), 93

\bibitem[Leinert et al.(1993)]{leinert1993} Leinert, C., Zinnecker, 
H., Weitzel, N., Christou, J., Ridgway, S.~T., Jameson, R., Haas, M., \& 
Lenzen, R.\ 1993, \aap, 278, 129 

\bibitem[Levato et al.(1987)]{levato1987} 
Levato, H., Malaroda, S., Morrell, N., \& Solivella, G.\ 1987, \apjs, 64, 487

\bibitem[Lindroos(1985)]{lindroos1985} Lindroos, K.~P.\ 1985, \aaps, 
60, 183 

\bibitem[Malkov(1993)]{malkov1993} Malkov, O.~Y.\ 1993, Bulletin 
d'Information du Centre de Donnees Stellaires, 42, 27 

\bibitem[Mamajek et al.(2002)]{mamajek2002}
Mamajek, E.~E., Meyer, M.~R., \& Liebert, J.\ 2002, \aj, 124, 1670 

\bibitem[Mart\'{\i}n et al.(2004)]{martin2004}
Mart\'{\i}n, E.~L., Delfosse, X., \& Guieu, S. \ 2004, Accepted for publication in \aj

\bibitem[Mason(1995)]{mason1995} Mason, B.~D.\ 1995, VizieR 
Online Data Catalog, 610, 70299 

\bibitem[Mathieu(1994)]{mathieu1994} Mathieu, R.~D.\ 1994, \araa, 
32, 465 

\bibitem[Mathis(1990)]{mathis1990} Mathis, J.~S.\ 1990, \araa, 28, 
37 

\bibitem[McAlister et al.(1993)]{mcalister1993} 
McAlister, H.~A., Mason, B.~D., Hartkopf, W.~I., \& Shara, M.~M.\ 1993, 
\aj, 106, 1639 


\bibitem[Miura et al.(1992)]{miura1992} Miura, N., Baba, N., 
Ni-Ino, M., Ohtsubo, J., Noguchi, M., \& Isobe, S.\ 1992, Publications of 
the National Astronomical Observatory of Japan, 2, 561 

\bibitem[Palla \& Stahler(1999)]{palla1999} Palla, F.~\& Stahler, 
S.~W.\ 1999, \apj, 525, 772 

\bibitem[Pedoussaut et al.(1996)]{pedoussaut1996} Pedoussaut, A., Capdeville, A., 
Ginestet, N., \& Carquillat, J.~M.\ 1996, VizieR Online Data Catalog, 4016, 

\bibitem[Portegies Zwart et al.(2001)]{spz2001} Portegies Zwart, S.~F., McMillan, 
S.~L.~W., Hut, P., \& Makino, J.\ 2001, \mnras, 321, 199 

\bibitem[Preibisch \& Zinnecker(1999)]{preibisch1999} Preibisch, 
T.~\& Zinnecker, H.\ 1999, \aj, 117, 2381 

\bibitem[Preibisch et al.(2002)]{preibisch2002} Preibisch, T., Brown, 
A.~G.~A., Bridges, T., Guenther, E., \& Zinnecker, H.\ 2002, \aj, 124, 404 

\bibitem[Preibisch, Stanke, \& Zinnecker(2003)]{preibisch2003} 
Preibisch, T., Stanke, T., \& Zinnecker, H.\ 2003, \aap, 409, 147 

\bibitem[Quist \& Lindegren(2000)]{quist2000}
Quist, C.F.~\& Lindegren, L.\ 2000, \aap, 361, 770

\bibitem[Reipurth \& Zinnecker(1993)]{reipurth1993} Reipurth, B.~\& 
Zinnecker, H.\ 1993, \aap, 278, 81 

\bibitem[Reipurth \& Clarke(2001)]{reipurth2001} Reipurth, B.~\& 
Clarke, C.\ 2001, \aj, 122, 432 

\bibitem[Sartori, L{\' e}pine, \& Dias(2003)]{sartori2003} Sartori, 
M.~J., L{\' e}pine, J.~R.~D., \& Dias, W.~S.\ 2003, \aap, 404, 913 

\bibitem[Savage \& Mathis(1979)]{savage1979} Savage, B.~D.~\& 
Mathis, J.~S.\ 1979, \araa, 17, 73 

\bibitem[Shatsky \& Tokovinin(2002)]{shatsky2002} Shatsky, N.~\& 
Tokovinin, A.\ 2002, \aap, 382, 92 

\bibitem[Sills et al.(2003)]{sills2003}
Sills, A.~et al.\ 2003, New Astronomy, 8, 605

\bibitem[Sowell \& Wilson(1993)]{sowell1993} Sowell, J.~R.~\& 
Wilson, J.~W.\ 1993, \pasp, 105, 36 

\bibitem[Svechnikov \& Bessonova(1984)]{svechnikov1984} Svechnikov, 
M.~A.~\& Bessonova, L.~A.\ 1984, Bulletin d'Information du Centre de 
Donnees Stellaires, 26, 99 

\bibitem[Testi, Palla, \& Natta(1998)]{testi1998}
Testi, L, Palla, F., \& Natta, A., 1998, \aaps, 133, 81

\bibitem[Tout(1991)]{tout1991}
Tout, C.~A., 1991, \mnras, 250, 701

\bibitem[Tokovinin(1997)]{tokovinin1997} Tokovinin, A.~A.\ 1997, 
\aaps, 124, 75 

\bibitem[Wolfire \& Cassinelli(1987)]{wolfire1987} Wolfire, 
M.~G.~\& Cassinelli, J.~P.\ 1987, \apj, 319, 850 

\bibitem[Worley \& Douglass(1997)]{worley1997} Worley, C.~E.~\& 
Douglass, G.~G.\ 1997, \aaps, 125, 523 

\bibitem[de Zeeuw \& Brand(1985)]{dezeeuw1985} de Zeeuw, T.~\& 
Brand, J.\ 1985, ASSL Vol.~120: Birth and Evolution of Massive Stars and 
Stellar Groups, 95 

\bibitem[de Zeeuw et al.(1999)]{dezeeuw1999} de Zeeuw, P.~T., 
Hoogerwerf, R., de Bruijne, J.~H.~J., Brown, A.~G.~A., \& Blaauw, A.\ 1999, 
\aj, 117, 354 

\end{thebibliography}
\end{document}